\documentclass[
reprint,
superscriptaddress,
amsmath,amssymb,
aps,
prx,
]{revtex4-1}

\usepackage[english]{babel}
\usepackage[utf8]{inputenc} 
\usepackage{physics}
\usepackage{ dsfont }
\usepackage{enumerate}
\usepackage{verbatim}
\usepackage[normalem]{ulem}
\usepackage{amssymb}
\usepackage{siunitx}
\usepackage{multirow}
\newcommand*\chem[1]{\ensuremath{\mathrm{#1}}}

\usepackage[colorlinks=false, linktocpage=true]{hyperref}

\usepackage{graphicx}
\usepackage{tikz}

\usepackage{dcolumn} 
\usepackage{bm}

\begin{document}

\title{Surface photogalvanic effect in Weyl semimetals}

\author{J.\ F.\ Steiner}
\affiliation{%
 Dahlem Center for Complex Quantum Systems and Fachbereich Physik, Freie Universität Berlin, 14195 Berlin, Germany
}%
\author{A.\ V.\ Andreev}
\affiliation{Skolkovo Institute of Science and Technology, Moscow, 143026, Russia}
\affiliation{Department of Physics, University of Washington, Seattle, Washington 98195, USA}
\affiliation{L. D. Landau Institute for Theoretical Physics, Moscow, 119334, Russia}
\author{M.\ Breitkreiz}
\affiliation{%
 Dahlem Center for Complex Quantum Systems and Fachbereich Physik, Freie Universität Berlin, 14195 Berlin, Germany
}%

\date{\today}

\begin{abstract}

The photogalvanic effect---a rectified current induced by
light irradiation---requires the intrinsic
symmetry of the medium to be 
sufficiently low, which 
strongly limits candidate materials for this effect.  
In this work we explore how in Weyl semimetals the photogalvanic effect can be 
enabled and controlled by design of the material surface.
Specifically, we provide a theory of  ballistic linear and 
circular  photogalvanic  current in a
Weyl semimetal spatially confined to a slab under general and
variable surface boundary conditions. The results 
are applicable 
to Weyl semimetals with an arbitrary number of Weyl nodes at radiation frequencies small compared to 
the energy of non-linear terms in the dispersion at the Fermi level. 
The confinement-induced response
is tightly linked to the configuration of Fermi-arc surface 
states, specifically the Fermi-arc connectivity and direction of emanation from the Weyl nodes, thus inheriting the same directionality and sensitivity to boundary conditions. As a result, the photogalvanic response of the system becomes much richer than that of an infinite system, and may be tuned via surface manipulations.

\end{abstract}

\maketitle

\section{Introduction}
In the past decade, Weyl semimetals (WSMs) have attracted great attention, from theoretical prediction \cite{wan_topological_2011} to experimental realization \cite{lv_experimental_2015,yang_weyl_2015,Xu2015b, Xu2015}.
Of particular interest are the peculiar transport phenomena
\cite{hosur_charge_2012,Burkov2017}
due to the presence of 
Weyl fermions, 
the associated chiral anomaly \cite{Adler1969,Bell1969, Nielsen1983}, and 
topological Fermi-arc surface states \cite{wan_topological_2011,Balents2011}. 
For instance, WSMs are 
considered a promising platform for optoelectronic applications \cite{liu_semimetals_nodate,Guan2021}
because chirality and the topologically protected linear dispersion of Weyl fermions generally tend
to enable and enhance the response to incident light
\cite{Armitage2017}. The relevant 
light frequencies lie typically in the mid- and far-infrared 
region, bounded from below by the typically small but finite chemical potential at the Weyl nodes and from above by the onsetting non-linear 
corrections to the Weyl dispersion. 

Most discussed is the photogalvanic effect (PGE) 
in non-centrosymmetric WSMs---a \textit{dc} current response to light irradiation \cite{taguchi_photovoltaic_2016,morimoto_semiclassical_2016-1,chan_photocurrents_2017,konig_photogalvanic_2017,golub_photocurrent_2017,sun_circular_2017,ma_direct_2017,zhang_photogalvanic_2018,PhysRevX.10.041041,Watanabe2021}. Generally, the photogalvanic current density may be expanded as \cite{belinicher_photogalvanic_1980,ivchenko2012superlattices}
\begin{equation}\label{eq:J} 
    \bm{J} = \sum_{i,j=x,y,z}\bm{\Gamma}_{ij} \mathcal{E}_i \mathcal{E}_j^*,
\end{equation}
where $\bm{\mathcal{E}}$ is the polarization vector of the light field and $\bm{\Gamma}$ is the photogalvanic response tensor. One distinguishes between a ballistic current, induced by asymmetric in momentum photogeneration (or injection following the terminology of \cite{PhysRevB.61.5337}), which is proportional to the relaxation time and dominates in clean samples, and the shift current, which is finite even in the absence of relaxation processes. 
Notably, in non-centrosymmetric WSMs a quantized photogeneration induced by circularly polarized light was predicted and observed \cite{de_juan_quantized_2017, rees_quantized_2019}. WSMs that in addition to inversion also break time-reversal symmetry may further exhibit a  ballistic response to linearly polarized light which may be giant \cite{chan_photocurrents_2017,Osterhoudt2019,Zhang2019a,PhysRevResearch.2.033100,Fei2020,Watanabe2021}.

\begin{figure}[b]
    \centering
    \includegraphics[width=\columnwidth]{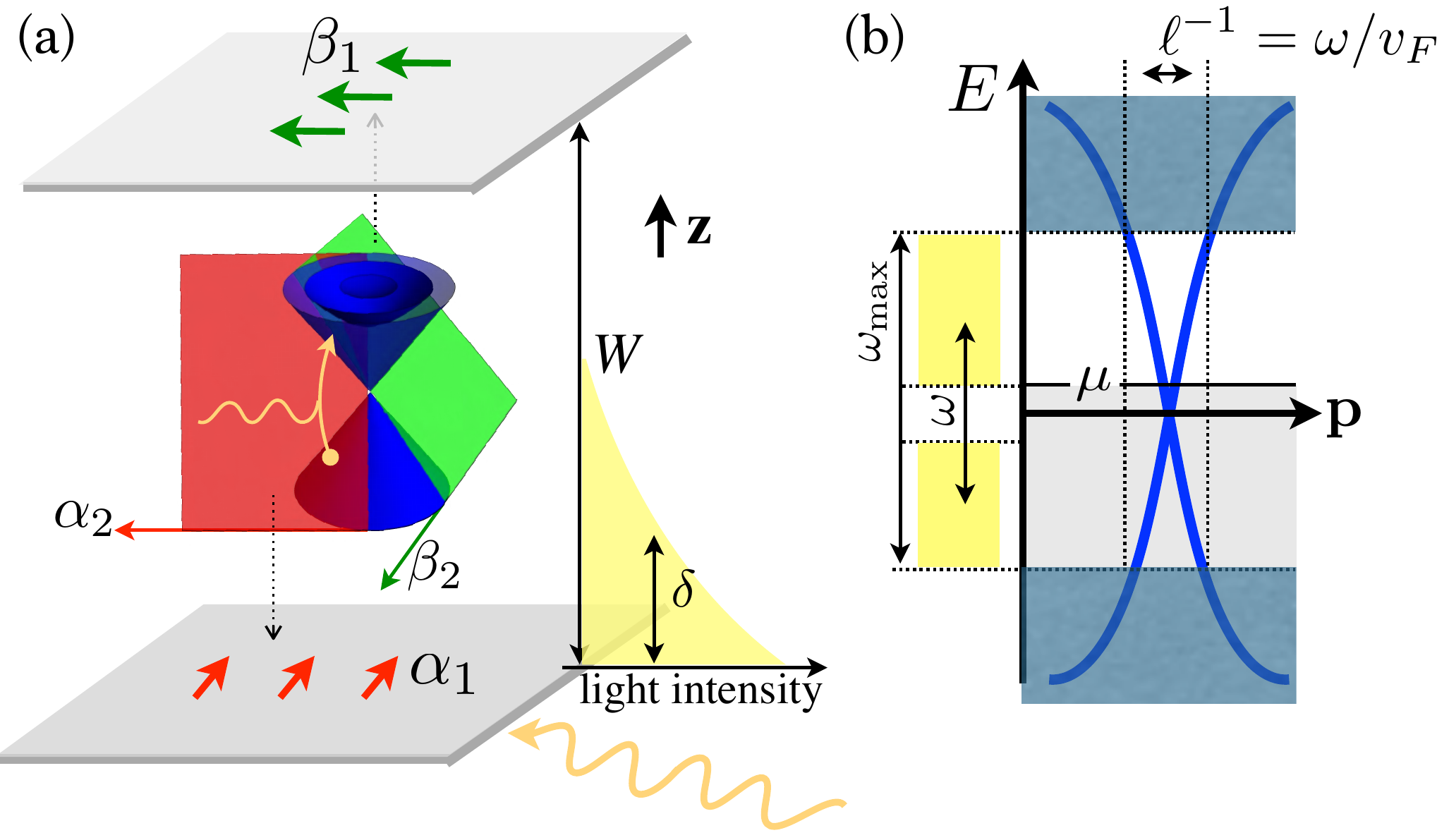}
    \caption{
    (a) Dispersion 
    (energy vs.\ in-plane momenta) 
    of Weyl fermions confined to a 
    slab of thickness $W$. The plot shows bulk states (blue)
    and surface states of bottom 
    surface (red) and top 
    surface (green). The surface states are tightly \textit{glued} to the bulk Weyl cone and emanate in the 
    direction $\bm{\alpha}_2$  ($-\bm{\beta}_2$) for bottom (top) surface. The figure also shows the
    directions of Fermi-arc motion
    ($\bm{\alpha}_1$ and $\bm{\beta}_1$), as well as the photon
    penetration depth $\delta$. (b) Low-energy
    band structure at the Weyl node. The Fermi 
    energy $\mu$ and the energy range of 
    the linear-dispersion regime, $\omega_\mathrm{max}$, determine 
    the range of  
    considered photon frequencies $2\mu<\omega<\omega_\mathrm{max}$.}
    \label{fig:basic_setup}
\end{figure}

Besides the bulk PGE that can be understood 
in terms of infinite-system models, the PGE has been explored
at the surfaces of metals 
\cite{Alperovich1981,MAGARILL1981,Moore2010} and topological insulators
\cite{Hosur2011,Junck2013,Lindner2015,PhysRevB.95.035134},
in which case the surface-normal component 
of $\bm{\Gamma}$ (but not that of $\bm{{\cal E}}$) vanishes.
In the field of WSMs, recently there was evidence from experiments and first principles calculations that Fermi arc states might play an 
important role in the photogalvanic response \cite{wang_robust_2019,chang_unconventional_2020}, a contribution that was neglected in previous theories. In
particular, \citet{chang_unconventional_2020} have shown that the 
contribution of surface states to the PGE due to excitations
between the surface states of the same surface 
are possible in some 
 crystals due to a non-linear dispersion of those surface states. 

The mere presence of surface states, however, does not capture the full 
peculiarity of a 
WSM. 
Importantly, the \emph{two}-dimensional Fermi-arc surface states, constituting in some sense
the reaction of a pair of chiral 
Weyl fermions to confinement, are tightly glued to 
the \emph{three}-dimensional Weyl fermions \cite{haldane2014attachment},
as illustrated in Fig.\ \ref{fig:basic_setup}(a). 
This \emph{connectivity} 
distinguishes Fermi arcs from  
surface states of metals and topological insulators and was shown to give rise to a number intriguing, counter-intuitive 
linear-response effects
\cite{Burkov2014,Moll2016,Wang2017b,Behrends2019,Zhang2019,Sukhachov2019, breitkreiz_large_2019,Kaladzhyan2019,
weylorbit2021,Breitkreiz2020,10.21468/SciPostPhys.11.2.046}.
Understanding its role also for the
photogalvanic response is highly desirable. The 
theoretical challenge  
to capture the 
effect of the connectivity is the requirement to go beyond 
an effective surface theory 
 and consider a full three-dimensional, yet 
spatially confined model.

In this work we present a theory of ballistic photogalvanic response of 
Weyl fermions spatially confined in one direction 
with general boundary conditions,
relevant for 
Weyl-semimetal slabs with an arbitrary configuration of 
Weyl nodes  \textit{and
arbitrary orientations  
of Fermi arcs at the bottom and top surfaces, which need not be the same.} 

Specifically, the orientation of the bottom (top) Fermi arc is defined by the direction of its velocity, $\bm{\alpha}_1$ ($\bm{\beta}_1$), 
 or the perpendicular direction at which the arc emanates from the Weyl node, $\bm{\alpha}_2$ (-$\bm{\beta}_2$), see Fig.\ \ref{fig:basic_setup}(a). 
 We show that this symmetry-breaking  directionality gives rise to a vastly richer response behavior compared to an 
unconfined WSM. In particular, 
the confinement enables   the otherwise vanishing linear 
and circular PGE in centrosymmetric WSMs. 
Furthermore, the response
is crucially determined by the orientations of the Fermi arcs. The latter may be
adjusted by choosing different surface terminations \cite{Morali2019, Fujii2021} or surface doping \cite{PhysRevB.92.201107}. 
In principle, 
this allows control over the photogalvanic response 
by modification of the surface only. 
 
To focus on Weyl physics, we consider a photonfrequency range for which excitations can take place only
close to Weyl nodes where the bulk and arc dispersions
are linear, see Fig. 1(b). The total response is then the
sum of the responses of individual Weyl nodes.
Further, we focus on the semimetallic 
regime, in which the Fermi level $\mu$ is close to the Weyl node and 
smaller than the photon energy, such that Pauli blocking as well as screening may be neglected.
In this regime, intra-surface (arc-arc) excitations are forbidden, but bulk-bulk excitations as well as arc-bulk excitations exist. 
 
Most strikingly, for a centrosymmetric WSM confined to a slab, 
the photogalvanic 
response  is fully determined by the Fermi-arc orientation.  
Considering the current density in Eq.~\eqref{eq:J} as 
 the current density averaged 
over the slab width, the response tensor
 can be decomposed into
a confinement-independent bulk-bulk contribution 
$\bm{\Gamma}^{\textrm{bb}}$ and confinement-induced
contributions, which in turn consist of bulk-bulk $\delta\bm{\Gamma}^{\textrm{bb}}$ as well as  
arc-bulk $\bm{\Gamma}^{\textrm{ab}}$ parts,
\begin{equation}\label{eq:contribs}
    \bm{\Gamma} = \bm{\Gamma}^{\textrm{bb}} +\delta \bm{\Gamma}^{\textrm{bb}}+  \bm{\Gamma}^{\textrm{ab}}. 
\end{equation}
For  a centrosymmetric WSM, $\bm{\Gamma^{\textrm{bb}}}$ vanishes according to
general symmetry considerations \cite{belinicher_photogalvanic_1980}.
The response is thus given by
\begin{equation}
    \bm{\Gamma}^{\textrm{centrosymm.}}= \delta\bm{\Gamma}^{\textrm{bb}} + \bm{\Gamma}^{\textrm{ab}},
\end{equation}
where both contributions are fully determined by the Fermi-arc orientations
since the orientation of the arcs and modification of the bulk-state wavefunctions are both defined by the boundary conditions.  Moreover, a centrosymmetric WSM necessarily breaks time-reversal symmetry, which implies that $\bm{\Gamma}^{\textrm{centrosymm.}}$ will include a ballistic response to linearly polarized light of the type discussed in \cite{Zhang2019a}. This is directly relevant to magnetic WSMs, such as \chem{Co_3Sn_2S_2} \cite{Liu2018a}, \chem{RhSi} \cite{rees_quantized_2019}, and \chem{GdPtBi} \cite{Suzuki2016a}.  Table \ref{tab:symmetry}
summarizes which types of photogalvanic response are possible in unconfined and confined 
WSMs, depending on the mechanism and the presence 
of time-reversal and inversion symmetry. 

\begin{table*} 
\begin{tabular}{ p{4cm} || p{3.3cm}|p{3.3cm}|p{5.5cm}  }
 symmetry & time reversal & inversion  & neither \\
\phantom{a}  &   (broken inversion)  & (broken time reversal) &  \phantom{a} \\
\hline
\hline
ballistic current  (injection)    & bCPGE, \textit{sCPGE} & \textit{sCPGE}, \textit{sLPGE} & bCPGE, bLPGE, \textit{sCPGE}, \textit{sLPGE}  \\
\hline
shift current  &  bLPGE, sLPGE & (sCPGE, sLPGE) & (bCPGE, bLPGE, sCPGE, sLPGE) \\
\end{tabular}
 \caption{
\label{tab:symmetry}
Allowed types of photogalvanic response in WSMs,
bCPGE, sCPGE, bLPGE, and sLPGE, distinguished by light polarization, circular (CPGE) and linear (LPGE),  
and origin (b for bulk) and (s for surface). Setups are 
categorized by mechanism (ballistic/shift current) and
presence/absence of time-reversal and inversion symmetry 
of the WSM material.
Terms in parentheses give subdominant response in clean systems. Italic text marks contributions first discussed in this work. 
In the presence of inversion symmetry, the bulk photogalvanic response vanishes and only surface contributions remain. In this case, the response is fully determined by the directionality of the Fermi arcs. In particular, there is a surface induced ballistic LPGE enabled by broken time-reversal symmetry.  }
\end{table*}

Finally, the confinement-induced PGE is categorized 
depending on the slab thickness. For a sufficiently thick slab or sufficiently high frequency the light field does not penetrate the whole slab. This is the case when the penetration depth $\delta$, which for photon energies
$\hbar\omega \sim $ \SI{1}{\milli\electronvolt} to \SI{1}{\electronvolt} lies in the 
range 
\SI{1}{\micro\metre} to \SI{1}{\milli\metre},  is much smaller than the slab thickness $W$. For light incident at the bottom surface, see Fig.\ \ref{fig:basic_setup}(a), the top surface no longer contributes to the response. This changes the symmetry of the response tensor. We refer to this limit as the \emph{thick slab}. In the opposite limit, referred to as the \emph{thin slab}, $\delta \gg W$, both surfaces contribute.
Technically, the two limits require substantially different
calculations, we will thus
mostly consider the thick- and thin-slab regimes separately, 
using different analytical and numerical techniques. 

This article is organized as follows.
In Sec.\ \ref{sec:model} we introduce the model of a WSM in the slab geometry for which we perform our calculations. We also briefly discuss the decay of light waves in WSMs. Finally, we present the semiclassical formulae for the photogalvanic current that we employ. 
In Sec.\ \ref{sec:classification} we classify the different contributions to the photogalvanic response tensor and estimate their magnitude. Further, we comment on the irrelevance of finite light momentum.
In Sec. \ref{sec:symmetry} we discuss the 
symmetry constraints on the response 
tensor. Finally, in Sec.\ \ref{sec:results} we present 
analytical results for the different contributions to the response tensor for a single Weyl cone 
in the different regimes.  We further present a lattice simulation in the thin limit which confirms the analytical results.
At the end of this section we 
apply our results to WSMs with several Weyl cones by considering a centrosymmetric WSM with two nodes. We conclude in Sec.\ \ref{sec:conclusion}. Technical details are delegated to the appendices.

\section{Model}\label{sec:model}

\subsection{Weyl semimetal}

We consider a WSM slab with a set of Weyl nodes which are close to the Fermi level and well-separated in momentum space. 
Since we consider the response to excitations
occurring close to 
the Weyl nodes only, it suffices to 
consider the response of a single Weyl node, from which the response 
of a WSM with several 
Weyl nodes will follow by 
combining the single-Weyl-node response tensors, transformed according to the specific Weyl-node 
arrangement. 

In order to evaluate the matrix elements relevant for the photogalvanic response tensors we seek explicit expressions for the wave-functions in the slab geometry 
(see Appendix A for a detailed derivation).
To this end, we model a single Weyl fermion confined to $0<z<W$ with the Hamiltonian
(we set $\hbar=1$)
\begin{align}\label{eq:ham_0}
    H =  \chi\,  v\, \bm{p}\cdot \bm{\sigma},  
\end{align}
where $\bm{p}$ is the momentum (with $p_z = -i\partial_z$), $\bm{\sigma}$ the 
spin, $\chi=\pm$ the chirality, and $v$ the  
velocity. For better transparency of the following calculations
we here assume isotropic velocity of 
the Weyl fermion; in Appendix \ref{app:anisotropic_weyl_node} we generalize the results
to an anisotropic Weyl node, which leads 
to a simple transformation of the 
response tensor. 
In the absence of a tilt, the Weyl Hamiltonian Eq. \ref{eq:ham_0} commutes with the operator ${\cal T} = i \sigma_y K$, where 
$K$ is complex conjugation. By analogy with relativistic theory we refer to this intra-node symmetry as time reversal (TR) symmetry. Note that it does not correspond to the time reversal operation acting on the whole crystal, as this connects different Weyl nodes. 
Thus the intra-node TR symmetry allows to constrain the response due to a single Weyl node only. A WSM with several Weyl nodes at generic points in momentum space clearly does not need to satisfy TR symmetry. 

Using translation invariance parallel to the surface we seek energy eigenstates in the form of plane waves 
in the $xy$ plane with the continuous in-plane momenta $\bm{p}_\parallel=(p_x, p_y) \equiv p_{\parallel} (\cos \phi, \sin \phi)$. Their dependence on $z$ is given by the solutions to the Weyl equation $H\psi(z)=E\psi(z)$, which may be written as 
\begin{align}\label{eq:wv0}
    \psi(z) \propto &\ \exp{i\mathcal{P}_z z} \psi(0) \nonumber\\ 
    \propto&\ \pqty{ p_z \cos(p_z z) + i \sin (p_z z) \mathcal{P}_z}\psi(0),
\end{align}
where $p_z= \sqrt{E^2-p_\parallel^2}$ and the generalized momentum operator reads
\begin{align}\label{eq:gen_mom}
     \mathcal{P}_z = \pqty{i p_y , -i p_x , \frac{\chi E}{v} }\cdot\bm{\sigma}.
\end{align}
The discrete energy eigenvalues of the slab 
(at fixed $\bm{p}_\parallel$)
are to be determined by boundary conditions.
A generic boundary condition
on the wavefunction 
is a vanishing
current $j_z$
across the boundaries. Since $j_z \propto\partial_{p_z}H\propto \sigma_z$ this corresponds to 
$\psi^\dagger\sigma_z\psi=0$. Accounting for the possibility of differing boundary conditions for the bottom and top surfaces, a general boundary condition thus reads
\begin{equation}\label{eq:bc}
    \psi(0) \propto \begin{pmatrix} 1 \\  e^{i\alpha} \end{pmatrix}, \;\;\;\;\;\ \ 
    \psi(W) \propto \begin{pmatrix} 1 \\  e^{i\beta} \end{pmatrix},
\end{equation}
parametrized by two independent angles $\alpha$ and $\beta$. Surface inhomogeneities would 
correspond to a spatial dependence of $\alpha$ and 
$\beta$. Here we assume  
translation invariance at the surface
(up to a relaxation mean free path that will be introduced perturbatively below) and thus consider
$\alpha$ and 
$\beta$ to be constant.

The boundary conditions lead to the equation
\begin{multline} \label{eq:bc_eq}
 \sin \frac{\beta-\alpha}{2} = \\  \frac{\tan (p_z W)}{p_z} \Bigg[  p_{\parallel}\cos\pqty{\phi-\frac{\beta+\alpha}{2}} \mp \chi p \cos \frac{\beta-\alpha}{2} \Bigg],
\end{multline}
which determines the quantized eigenvalues $p_z$. 
Solutions with real $p_z$ correspond to bulk states, imaginary solutions correspond to surface ``arc" states. For details and explicit expressions of the arc and bulk states see App. \ref{app:wave-functions}. Note that $\alpha$ and $\beta$ define the velocity of the Fermi arcs localized at the bottom (b) and top (t) surfaces, 
\begin{align}
    \bm{v}^{\mathrm{b}}_{\textrm{arc}} =&\ \chi v 
    \bm{\alpha}_1,\ 
    \bm{v}^{\mathrm{t}}_{\textrm{arc}} = \chi v \bm{\beta}_1.
\end{align}
as well as the direction at which they emanate from the Weyl node, given by the constraint
\begin{align}\label{eq:arcconstr}
      \bm{p}\cdot\bm{\alpha}_2 \equiv \kappa_\mathrm{b} >0  ,\ \ \ \  \bm{p}\cdot(-\bm{\beta}_2) \equiv \kappa_\mathrm{t} > 0  ,
\end{align}
for bottom and top arc, respectively, where we defined the vectors
\begin{subequations}
\begin{align}
    \bm{\alpha}_1 =&\ \begin{pmatrix} \cos \alpha  \\  \sin \alpha \\ 0 \end{pmatrix},\ 
    \bm{\alpha}_2 = \begin{pmatrix} -\sin \alpha  \\  \cos \alpha \\ 0  \end{pmatrix}, \\
    \bm{\beta}_1 =&\ \begin{pmatrix} \cos \beta  \\  \sin \beta \\ 0 \end{pmatrix},\
    \bm{\beta}_2 = \begin{pmatrix} -\sin \beta  \\  \cos \beta \\ 0  \end{pmatrix}.
\end{align}
\end{subequations}
The quantities $\kappa_\mathrm{t}$ and $\kappa_\mathrm{b}$ introduced in Eq.~\eqref{eq:arcconstr} have the meaning of inverse decay lengths of the evanescent wave functions of  arc states at the top and bottom surfaces respectively. 

In order to analyze symmetries in the presence of the boundary conditions, it proves helpful to define an equivalent multilayer setup, which reproduces the same spectrum and wave-functions as the boundary conditions Eq.~\eqref{eq:bc}. Note that this is a fictitious system only introduced to assist in understanding the response of a single Weyl node. The equivalent multilayer setup is defined by the Hamiltonian
\begin{equation}\label{eq:Hhetero}
    H_{\chi m } =  \chi v \bm{p}
\cdot \bm{\sigma} +\begin{cases}
- \chi m\, \bm{\sigma}\cdot\bm{\alpha}_2 & z<0 \\ 0 & 0<z<W \\
 \chi m\, \bm{\sigma}\cdot\bm{\beta}_2 & z>W
\end{cases},
\end{equation}
with $m \to\infty$ \cite{berry1987,bovenzi_twisted_2018}. Under TR the 
multilayer Hamiltonian transforms like
\begin{equation}\label{eq:TRtrafo}
  {\cal T}^{-1}H_{\chi m}{\cal T} = H_{\chi -m}.
\end{equation}
The mass terms of the boundary 
conditions thus behave like TR-breaking magnetizations 
in the directions $- \chi   \bm{\alpha}_2$ and 
$\chi   \bm{\beta}_2$ at the two boundaries. Note that this does not imply TR-breaking of the WSM with several Weyl nodes.

Furthermore, note that the directions of the boundary spinors can be additionally controlled by  
TR-\emph{preserving} boundary potentials \cite{PhysRevB.92.201107}. One 
can easily check that adding
 a boundary 
potential $\delta H_\mathrm{b} = 
\delta(z) \mu_0 + \delta(z-W)\mu_W$ to the Hamiltonian 
\eqref{eq:Hhetero}, 
rotates the boundary spinors 
like $\alpha\to\alpha+\chi 2\mu_0$ and 
$\beta\to\beta-\chi 2\mu_W$.
Boundary potentials are typically disregarded in minimal models 
of Weyl-semimetal slabs, which corresponds
to straight arcs connecting the Weyl cones, i.e., 
$\beta = \alpha + \pi$. Here we instead consider the general case that 
the Fermi arcs can emanate in any direction, 
considering
the boundary spinors \eqref{eq:bc} to be
given by  two independent variables $\alpha$ and $\beta$.
The resulting 
curvature of Fermi arcs, which is necessary to connect pairs of Weyl nodes and is 
often observed in experiments, is irrelevant in the close 
vicinity of the 
Weyl nodes to which the optical transitions that we consider are bound.

\begin{figure}
    \centering
    \includegraphics[width=0.9\columnwidth]{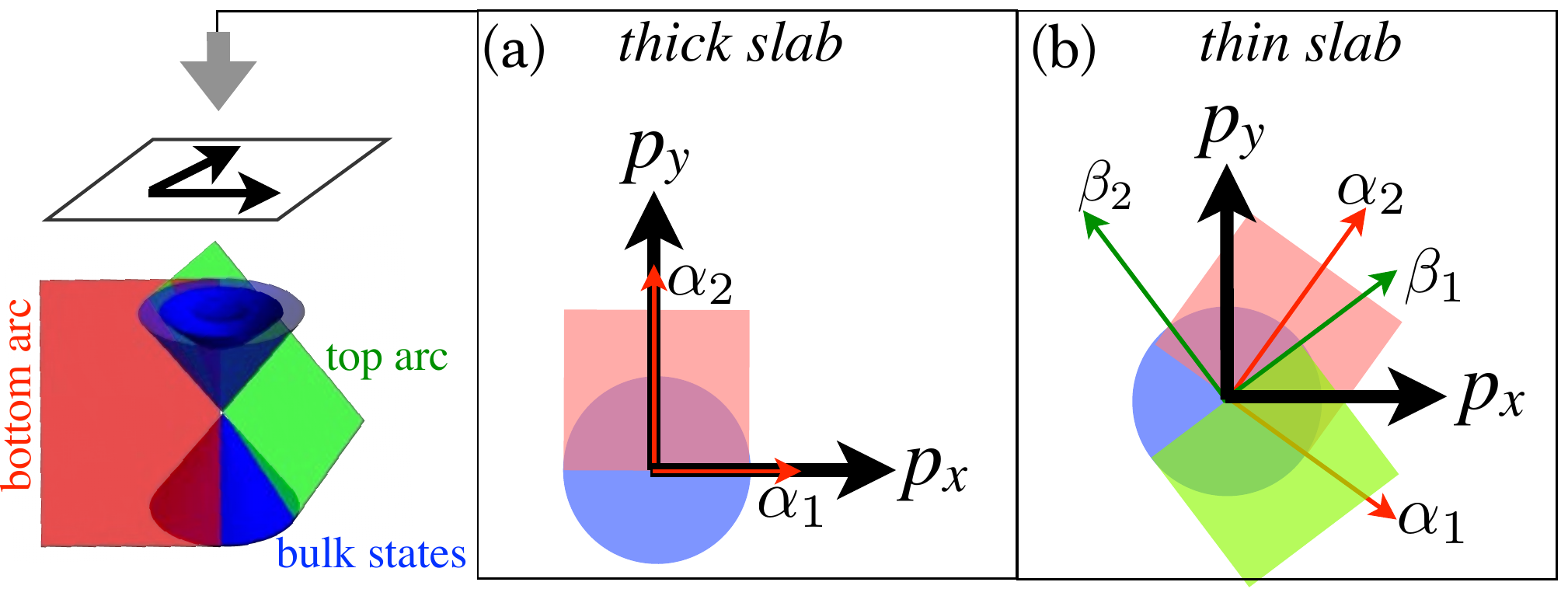}
    \caption{Top view on the slab dispersion 
    (left, see also Fig.\ \ref{fig:basic_setup}) showing the
    choice of coordinate axes 
    $\bm{p}_x$ and $\bm{p}_y$ to lie along
    high-symmetry directions in (a) the thick-slab ($W\gg\delta$)
     and (b) the thin-slab case
     ($W\ll\delta$). The slab dispersion
      features bulk states (blue), surface states of bottom (red) and top (green) surfaces.
     In (a) only the bottom surface matters since the light penetrating over the depth $\delta$ at the bottom surface does not reach the top arc.    }
    \label{fig:directions}
\end{figure}
The directionality 
introduced by the 
boundary conditions
will crucially determine the direction of 
the response. It is therefore
convenient to define the coordinate axes along the emergent 
high-symmetry directions. Those 
depend on whether current is induced 
at a single surface 
(thick-slab case)
or in the whole slab (thin-slab case). 
Figure \ref{fig:directions} illustrates
the geometry and the high-symmetry axes in these 
two cases.

\subsection{Electromagnetic waves in Weyl semimetals}\label{sec:wsm_em_waves}

For frequencies $\omega \gtrsim \mu$ the conductivity in WSMs is given by \cite{hosur_charge_2012,steiner_anomalous_2017}
\begin{equation}
    \sigma(\omega) = N \frac{e^2}{24\pi \epsilon_{\infty} v} \abs{\omega} = N \frac{\xi}{24\pi} \abs{\omega},
\end{equation}
Here, $N$ is the total number of Weyl nodes in the system, $\epsilon_{\infty} \sim 10$ is the permittivity due to inert bands and we let $e^2 \to e^2 / \epsilon_{\infty}$ to account for screening. Finally, we defined the dimensionless coupling constant 
\begin{align}
    \xi = \frac{e^2}{  \epsilon_{\infty} v} = \frac{1}{137} \times \frac{c}{v} \times \frac{1}{\epsilon_{\infty}} \sim 0.1.
\end{align} 
Note that ${\cal N} = N \xi /6$  
takes values between $1/30$ and $1$ in a WSM, depending 
on the number of nodes. 
The imaginary part of $\sigma$ has only weak frequency dependence and has been absorbed into $\epsilon_{\infty}$.
The frequency dependent permittivity then reads
\begin{equation}
    \epsilon(\omega) = \epsilon_{\infty}\bqty{1 + \frac{4\pi i}{\omega} \sigma(\omega) } = \epsilon_{\infty}\bqty{1 + i \frac{N\xi}{6} \textrm{sign}(\omega) }. 
\end{equation}
We consider light entering the WSM at the $z=0$ surface. The field \textit{inside} the WSM has the form,
\begin{equation}
    \bm{E}(\bm{r},t) \propto \exp{i \pqty{\bm{k} \cdot\bm{r}  - \omega t}} \exp{- z/\delta},
\end{equation}
where $\bm{k}$ is the momentum inside the medium and $\delta$ is the penetration depth. In terms of the vacuum wavenumber $k_v = \omega/c$ and to leading order in $\mathcal{N}$, they are given by
\begin{subequations} \label{eq:field_momentum_decay}
\begin{align}
    \abs{\bm{k}} =&\ \sqrt{\epsilon_{\infty} }k_v \pqty{1 + {\cal N}^2}^{1/4} \cos{\frac{\arctan {\cal N}}{2}} \simeq \sqrt{\epsilon_{\infty} } k_v, \\
    \frac{1}{\delta} =&\ \sqrt{\epsilon_{\infty}} k_v \pqty{1 + {\cal N}^2}^{1/4} \sin{\frac{\arctan {\cal N}}{2}} \simeq \frac{\sqrt{\epsilon_{\infty} } {\cal N}}{2}  k_v.
\end{align} 
\end{subequations}
With the above estimate of ${\cal N}$, depending on the number of Weyl nodes, we thus obtain $k\delta \sim 1 \dots 10$.

\subsection{Photogalvanic response tensor}\label{sec:pg_response_tensor}

We consider the response of the Weyl slab
to the weak external oscillating electric field 
\begin{equation}
    \bm{E}(\bm{r},t) = \bqty{\bm{\mathcal{E}} e^{i\bm{k}\cdot\bm{r}-i\omega t} + \textrm{c.c.}}e^{-z/\delta}.
\end{equation}
In the temporal gauge $\bm{E} = -\partial_t \bm{A}$, the perturbation to the Hamiltonian reads
\begin{align}
     \delta H  = \bm{j} \cdot \bm{A}(\bm{r},t) =&\  i \chi e \ell  \bm{\mathcal{E}} \cdot \bm{\sigma} e^{-i\omega t}e^{i\bm{k}\cdot\bm{r}- z /\delta } + \textrm{h.c.},
\end{align}
where $\bm{j} = -e \chi v \bm{\sigma} $ is the 
current operator and  $\ell = v/\omega$ is the smallest length scale of our model.
In the following we 
will use dimensionless length and momenta, denoted with a tilde, 
\begin{equation}
    \tilde{\bm{r}} = \frac{\bm{r}}{\ell},
    \ \ \ \ \ 
    \tilde{\bm{p}} = \bm{p}\, \ell,
\end{equation}
in units of $\ell$ and $\ell^{-1}$, respectively.

The ballistic PGE can be described within the framework of the Boltzmann kinetic equation by balancing asymmetric photogeneration and impurity-induced relaxation. Using the standard perturbation theory and relaxation-time approximation, one can express the photogalvanic response in terms of the momentum relaxation time $\tau$ in the form 
\cite{chan_photocurrents_2017}
\begin{align}\label{eq:pge_formula_main}
    \bm{\Gamma}_{ij} =&\  \frac{8\pi\eta \tau}{  \tilde{W}} 
    \int d^2\tilde{p}_{\parallel}\, \sum_{q_z p_z }  
    \pqty{\frac{\bm{v}_{\bm{p}+}}{v}  
     -\frac{ \bm{v}_{\bm{q}-}}{v} } \nonumber    \\
&\ \ \times \delta 
\pqty{1 - \frac{E_{\bm{p}}}{ \omega }
-\frac{E_{\bm{q}}}{ \omega}}
\big(\bm{M}_{\bm{p}\bm{q}} \otimes \bm{M}_{\bm{p}\bm{q}}^*\big)_{ij}
\end{align}
where $\bm{p} = (\bm{p}_\parallel,p_z)$, 
$\bm{q} = (\bm{p}_\parallel -\bm{k}_\parallel,q_z)$, $d^2 \tilde{p}_\parallel = \ell^2 d\bm{p}_\parallel$, and we introduced the matrix 
elements
\begin{align} \label{eq:matrixelements}
     \bm{M}_{\bm{p}\bm{q}} =&\, \bra{+, \bm{p}} \bm{\sigma} e^{i\bm{k}\cdot\bm{r}- z/\delta} \ket{-, \bm{q}},
\end{align} 
and the constant (restoring $\hbar$, which is set to one)
\begin{equation}
 \eta = \frac{e^3}{16\pi^2 \hbar^2}.
\end{equation}
These expressions hold for both bulk-bulk and and arc-bulk excitations. To avoid overcounting of states, for bulk states the sum runs only over $p_z >0$ while for arc states it runs over $\Im{p_z} > 0$. 
Note that $\Gamma_z = 0$ as $v_z = 0$ for all states due to the boundary conditions. 

 Note that the three $3\times 3$ matrices 
 $\bm{\Gamma}$ are hermitian. According to 
 standard terminology, the imaginary anti-symmetric part  is associated with  the circular PGE, which is present  only  
 if the incident light is elliptically polarized (the inverse implication is not true: elliptically polarized light 
 can give rise to photogalvanic response
 stemming from the real symmetric part). 
 The real symmetric part is referred to as the linear photogalvanic response, which exists even for linearly polarized radiation.

\section{Classification and estimate of
response contributions}\label{sec:classification}

There are three relevant length scales
in the problem 
 \footnote{Here we neglect one length scale of the problem, which is the 
mean free path $\tau v$ given by the relaxation time $\tau$.
Within the semiclassical approach described in 
Sec.~\ref{sec:pg_response_tensor} the mean free path is 
assumed long compared to essentially all other relevant scales,
which makes the mean free path itself irrelevant for the 
following discussion.}, 
the $v/c$-weighted light 
wavelength $\ell=v/\omega$, the light 
penetration depth
$\delta$,  
and the slab thickness $W$, whereby the 
weighted light wavelength is always much smaller than
the penetration depth,  $\ell/ \delta \sim v/c \sim 10^{-2}$.
The width $W$ is considered in two limits, 
the \emph{thick-slab case} $W \gg \delta$ and the 
 \emph{thin-slab case}  $\delta \gg W $. 
 In the thick-slab case the light completely decays inside the
 slab and only 
 a single slab surface is excited.
 In the thin-slab case the light 
 penetrates nearly homogeneously the whole slab such that 
 both surfaces are equally excited.
 In this limit, for simplicity
 of analytical calculations  
 we introduce a lower bound for the width, $W\gg \ell$, so 
 that energy quantization of slab modes is small compared 
 to the light frequency. The 
 ultrathin  case $W\sim \ell$ will be considered numerically 
 on a lattice model. 

Before coming to the detailed calculation, it 
is useful to classify the  
response contributions according to their 
dependencies on the relevant length scales 
($\ell$, $\delta$, $W$), 
separating confinement-independent from confinement-induced contributions and distinguishing contributions due to
arc-bulk and bulk-bulk excitations as 
given in \eqref{eq:contribs}. 
The result is summarized in Table \ref{tab:length}
and is explained in the following. 

\begin{table} 
\begin{tabular}{ p{2.2cm}||p{2.8cm}|p{2.8cm}  }
 & thick slab $W \gg \delta$ & thin slab $\delta \gg W$ \\
 \hline
 \hline
$\bm{\Gamma}^\mathrm{bb} $   &  
$\frac{\delta}{W} = \frac{\tilde{\delta}}{\tilde{W}}$	    &  $1$ \\
\hline
$\delta \bm{\Gamma}^\mathrm{bb}$, $\bm{\Gamma}^\mathrm{ab}$   &  
$\frac{\ell}{W} = 
\frac{1}{\tilde{W}}$& 
$\frac{\ell}{W} = \frac{1}{\tilde{W}}$ 
\end{tabular}
\caption{Scaling of the confinement-independent 
contribution $\bm{\Gamma}^\mathrm{bb}$ and the 
confinement-induced contributions 
$\delta \bm{\Gamma}^\mathrm{bb}, 
\bm{\Gamma}^\mathrm{ab}$
 with relevant length scales of the
system ($W$, $\ell$, $\delta$) 
in the cases of a thin and thick slab. }
\label{tab:length}
\end{table}

To estimate the magnitudes of contributions it suffices to disregarding the spin degree of freedom and 
consider the bulk wavefunctions to be of the form 
$\ket{\bm{q}} = \exp(i q_z z)/\sqrt{W}$ and that of arc
states of the form $\ket{\bm{q}} = \exp(-z/\ell)/\sqrt{l}$. 
In the latter, the  inverse decay length
$\kappa$, given in \eqref{eq:arcconstr}, 
has been approximated by 
the typical inverse distance from the Weyl node in the 
active region of excitations, which is 
set by $\ell^{-1}\equiv \omega/v$. Neglecting the in-plane light momentum
$\bm{k}$ (will be justified below), 
the  matrix elements \eqref{eq:matrixelements} for the thick-slab case can be estimated as
\begin{equation}\label{eq:matrixel}
|\bm{M}|^2 \sim
\begin{cases}
 \pqty{\frac{\delta}{W}}^2 \frac{1}{1+\bqty{\pqty{q_z-p_z}\delta}^2} & \text{bulk-bulk}\\ 
 \frac{\ell}{W}\frac{1}{1+\pqty{q_z\ell}^2} &
\text{arc-bulk}.
\end{cases}
\end{equation}
The momentum separation of modes 
is $1/W$, hence the number of modes within the active range
 around the node is $W/\ell$. 
The summation over 
$p_z$ and $q_z$ thus gives
\begin{equation}\label{eq:matrixelsum}
\sum_{p_zq_z}|\bm{M}|^2 \sim 
\begin{cases}
\frac{\delta}{\ell}  & \text{bulk-bulk}\\
1&
\text{arc-bulk}
\end{cases}
\end{equation}
 and the magnitude of the 
 response tensor will thus scale like
\begin{equation}\label{eq:scales_J}
\Gamma^\mathrm{bb} \sim  
\frac{\delta}{W},  \ \ \ \ \ \  
\Gamma^\mathrm{ab} \sim  
\frac{\ell}{W},
\end{equation}
for bulk-bulk and arc-bulk excitations, 
respectively.

Since $\delta\gg \ell$, bulk-bulk excitations will
give the 
dominant current contribution, while the confinement-induced correction due to arc-bulk
excitations give the finite-size correction 
with the small parameter $\ell/\delta$. 
Importantly, there are also contributions due to 
bulk-bulk excitations possible that scale like those from
arc-bulk excitations,
\begin{equation}
 \delta \Gamma^\mathrm{bb} \sim 
 \Gamma^\mathrm{ab}.
\end{equation}
To see this, note that the contribution 
$\Gamma^\mathrm{bb} $ stems
from approximating the peaked behavior of the 
bulk-bulk matrix elements  in \eqref{eq:matrixel}  at 
$q_z=p_z$ by a delta function,  
the correction to setting $q_z=p_z$ is of the order 
$\ell/\delta$ because the peak width is 
$1/\delta$ and the 
effective integration  range  $1/\ell$. 
Hence the leading correction scales like the arc-bulk contribution,
and needs to be taken into account.

Upon changing the scales from the thick-slab case, 
$W\gg\delta$, to 
the thin-slab case, $\delta\gg W$, the scaling of the contribution of arc-bulk excitations does not change
because the localization length of most arc states, 
$\kappa^{-1}$ given in \eqref{eq:arcconstr}, is set by 
$\ell$ and hence much smaller than both $W$ and $\delta$. 

For bulk-bulk excitations, the matrix elements are now the 
overlaps of wavefunctions over the whole slab width, 
\begin{equation}\label{eq:matrixel2}
|\bm{M}|^2 \sim
 \frac{1-\cos\bqty{\pqty{q_m-q_n}W}}{\bqty{\pqty{q_m-q_n}W}^2}.
\end{equation}
Summation over 
$p_z$ and $q_z$ gives 
$
\sum_{p_zq_z}|\bm{M}|^2 \sim 
\tilde{W}
$
and the magnitude of the 
 current thus scales like
\begin{equation}
\Gamma^\mathrm{bb} \sim 1,
\end{equation}
missing the factor $\delta/W$ as compared to the 
limit $W\gg \delta$ given in \eqref{eq:scales_J}, since transitions are now produced across the full width of the 
slab.

As before, the matrix elements are peaked at  $q_m=q_n$; the correction $\delta \Gamma^\mathrm{bb}$ to the $q_m=q_n$ contribution $ \Gamma^\mathrm{bb}$ 
is of order $l/W$
because the peak width is now $1/W$, while the integration 
range is still $1/\ell$. Thus $\delta \Gamma^\mathrm{bb}\sim
 \Gamma^\mathrm{ab}$ remains valid also in the thin-slab limit.  This concludes the explaination of the 
 scaling summarized in Table \ref{tab:length}.

\subsection{Irrelevance of the light momentum}

The momentum transfer due to a finite light momentum has 
the magnitude $k \sim \omega/c$. The small parameter of 
corrections due to this momentum shift is $k/p$, where
$p \sim 1/\ell = \omega/v$ is the typical 
momentum of excited states, hence
$k/p \sim v/c \sim 0.01$.
Comparing the smallness of corrections to the response, those
due to a finite
$\bm{k}$ are irrelevant for the thin-slab case 
but potentially relevant
in the case of a thick slab, where they are on the 
same order as the finite-size corrections, cf.\ Table \ref{tab:length}.
It turns out, however, that corrections 
to leading order in $k/p\sim v/c$ vanish also for the thick-slab case,  which we show 
explicitly for our slab model in Appendix
\ref{app:calculation_bb}. 
An easier way to find the same result is to realize that
 considering 
the correction due to a finite $k/p$, one can  neglect the 
finite-size corrections, which would give terms that are quadratic
in the small parameter.  Neglecting finite-size corrections, the 
result should thus coincide with that of an infinite system. 
In particular, the directionality introduced by the confinement 
becomes irrelevant. It is straightforward to 
verify that for a bulk Weyl cone the first-order 
$k/p$ corrections vanish.

For the  response tensor in 
 Eq.\ \eqref{eq:pge_formula_main}
this means that $\bm{k}$ can be set to zero, 
the matrix elements become
\begin{equation} \label{eq:matelnew}
\bm{M}_{\bm{p}\bm{q}} = \bra{+, \bm{p}} \bm{\sigma} e^{- z /\delta} \ket{-, \bm{q}},
\end{equation}
and the momenta have the same parallel component,  $\bm{p} = (\bm{p}_\parallel,p_z)$,   
$\bm{q} = (\bm{p}_\parallel,q_z)$.

\section{Symmetry constraints}\label{sec:symmetry}

As the last preliminary consideration
before coming to the explicit 
results, we now consider 
symmetry constraints on the response 
tensor. Considering the transition matrix elements \eqref{eq:matelnew}
we realize that since the band
index $\pm$ enters the  
wavefunctions in the form
$\pm\chi$, which can be explicitly seen in 
Appendix \ref{app:wave-functions}, 
Eq.\ \eqref{eq:bwv},
we obtain the relation
\begin{equation}
    \bm{M}_{\bm{p}\bm{q}}^*
    \big|_\chi
    =\bm{M}_{\bm{q}\bm{p}}\big|_{-\chi}.
\end{equation}
Using this and that 
other terms in the 
response expression, Eq.\ 
\eqref{eq:pge_formula_main},
are symmetric in 
$\bm{p}\leftrightarrow\bm{q}$,
we conclude that 
\begin{equation}
    \pqty{\bm{\Gamma}\pm \bm{\Gamma}^T}_\chi
    =\pm \pqty{\bm{\Gamma}\pm \bm{\Gamma}^T}_{-\chi},
\end{equation}
showing that the (anti)symmetric
part of the response tensor is 
even (odd) in the
chirality $\chi$. Moreover, generally
the (anti)symmetric
part of the response tensor is odd (even)
under TR \cite{belinicher_photogalvanic_1980}, which, according to the transformation behavior
\eqref{eq:TRtrafo} is given by 
$m\to- m$ (in the fictitious multilayer system) and thus
\begin{equation}
    \pqty{\bm{\Gamma}\pm \bm{\Gamma}^T}_m
    =\mp \pqty{\bm{\Gamma}\pm \bm{\Gamma}^T}_{-m}.
\end{equation}
For the thick-slab case, only the bottom surface is involved and 
$m\to -m$ corresponds to inversion of
$\bm{\alpha}_2
\equiv \hat{\bm{y}}$, i.e., 
mirror reflection $R_y$ with
respect to the $xz$ plane.
Taking into account also symmetry with respect to $R_x$,
the response tensor assumes the form 
\begin{multline}
     \bm{\Gamma}^\mathrm{thick}  
=\ \hat{\bm{x}}\begin{pmatrix}
0& \Gamma_{xxy}  & 0 \\
\Gamma_{xxy} & 0&  \Gamma_{xyz} \\
0& -\Gamma_{xyz} & 0 \end{pmatrix}\\
+\hat{\bm{y}} \begin{pmatrix}
\Gamma_{yxx} & 0  & - \Gamma_{yzx} \\
0 & \Gamma_{yyy} & 0 \\
\Gamma_{yzx}& 0 & \Gamma_{yzz} \end{pmatrix}.
\end{multline}

For the thin-slab case, both surfaces are involved and 
combinations of two reflections
leave the Hamiltonian invariant 
or time-reversed. In the
thin-slab basis [Fig.\ \ref{fig:directions}(c)] we obtain 
\begin{subequations}
\begin{align}
       R_yR_z  H_{\chi m } R_zR_y =&\  H_{\chi m },\\
    R_xR_y H_{\chi m } R_y R_x =&   H_{\chi -m}.
\end{align}
\end{subequations}
The resulting transformation 
behavior of the response tensor
dictates the form
\begin{multline}
     \bm{\Gamma}^\mathrm{thin}  
=\ \hat{\bm{x}}\begin{pmatrix}
\Gamma_{xxx} & 0  & 0 \\
0&  \Gamma_{xyy} &  \Gamma_{xyz} \\
0& -\Gamma_{xyz} & \Gamma_{xzz} \end{pmatrix}\\
+\hat{\bm{y}} \begin{pmatrix}
0 & \Gamma_{yxy}  & - \Gamma_{yzx} \\
 \Gamma_{yxy} & 0 & 0 \\
\Gamma_{yzx}& 0 & 0 \end{pmatrix}.
\end{multline}
A more detailed derivation of the tensor forms is given in  Appendix \ref{app:symmetry}.

\section{Results}\label{sec:results}
\subsection{PGE due to arc-bulk excitations}
\label{sec:arc-bulk}

\begin{figure}
\includegraphics[width=\columnwidth]{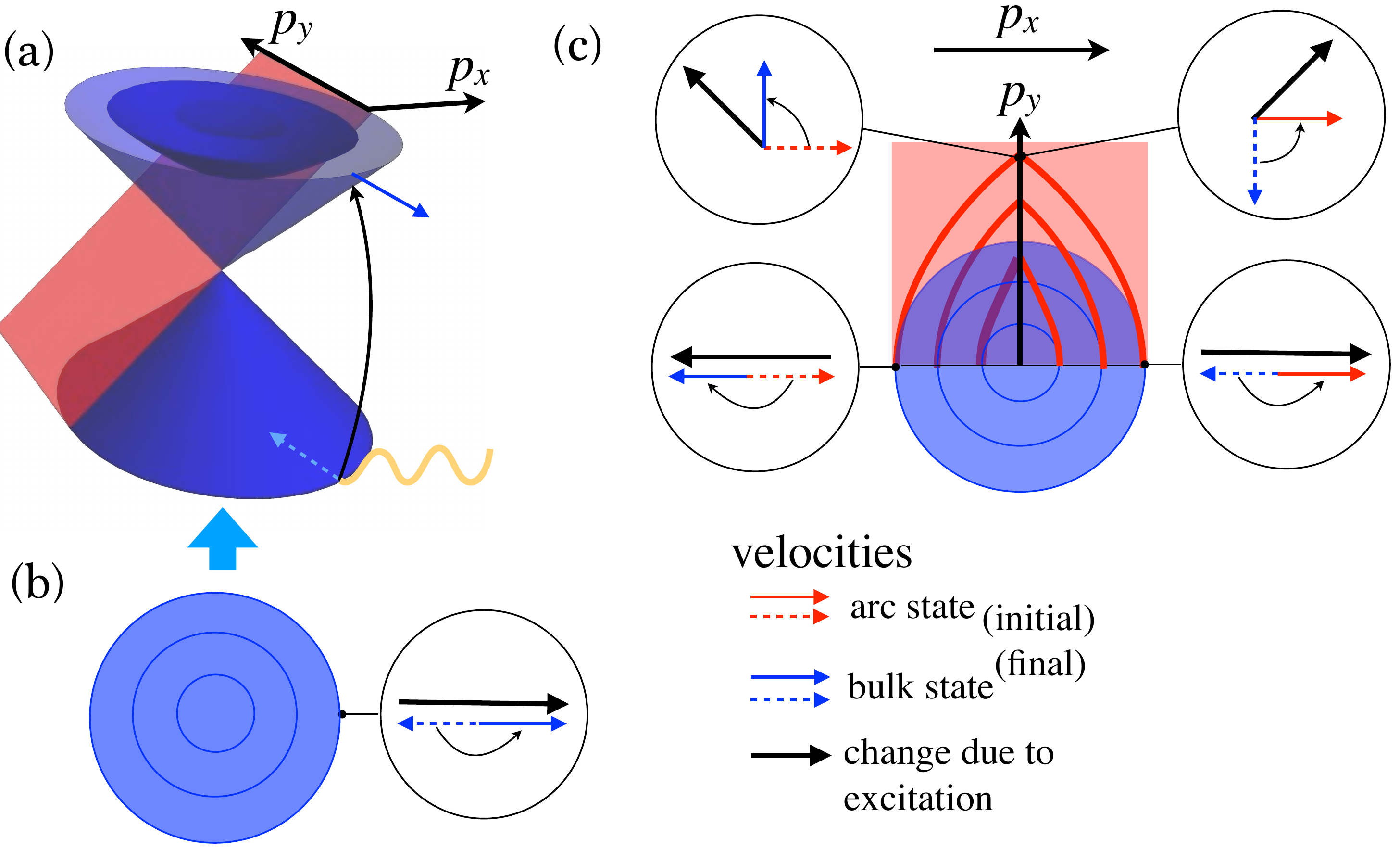}
\caption{(a) Dispersion of Weyl fermions 
confined to a slab as function of in-plane momenta in the thick slab basis.
Bulk states are colored blue and the  Fermi-arc surface states red
(only the bottom surface is shown). Velocities of 
initial (dashed arrow) and final state (solid arrow) 
of a photoexcitation are indicated. (b) Top view on 
bulk states in (a) showing the 
velocity change due to a bulk-bulk excitation. 
(c) Same as (b) but with indicated velocity change 
from  arc-bulk excitations. Red contours 
indicate those states of the surface arc states   that
satisfy the energy-conservation constraint for 
arc-bulk excitations $ \omega = \chi v p_x \pm v p$.}
\label{fig:arc_bulk_transitions}
\end{figure}

Arc-bulk excitations give rise to a current
that is ``automatically'' a 
 finite-size effect. Other finite-size 
corrections are negligible, which can be used to simplify 
the expression for the response tensor in Eq.\ \eqref{eq:pge_formula_main};
we can disregard the quantization of modes and
replace the sums by integrals. The integration over $z$ in the 
matrix elements of Eq.\ \eqref{eq:matelnew} may be extended to 
infinity since the decay of surface modes at most momenta 
is on the order of $\ell\ll \delta,W$, in both the 
thick-slab and thin-slab limits. Moreover we can neglect
confinement-induced corrections 
of bulk states.
A straightforward calculation (see Appendix \ref{app:calculation_ab} for details)
then gives 
\begin{align}\label{eq:resultab}
    \bm{\Gamma}^{\textrm{ab,thick}}_{ij}  =&\ \frac{2\pi\eta\tau}{\tilde{W}}\  
    \bqty{i\chi \frac{8}{3}\varepsilon_{xij} \hat{\bm{x}}
    +\ln (2)\,  \delta_{ij}(1-\delta_{xi})\hat{\bm{y}}}
\end{align} 
for the bottom arc in the thick-slab basis
$\hat{\bm{x}} =\bm{\alpha}_1 $, $\hat{\bm{y}} =\bm{\alpha}_2 $
[Fig.\ \ref{fig:directions}(b)].
The antisymmetric part is expressed using the
Levi-Civita symbol $\varepsilon_{ijk}$. This is the only arc-bulk contribution in the thick-slab case.

In the thin-slab case we add the contribution of the 
top arc, which is equivalent to the bottom arc up to the 
changed directions, $\bm{\alpha}_1\to \bm{\beta}_1$,
$\bm{\alpha}_2\to -\bm{\beta}_2$, see Fig.\ \ref{fig:directions}. 
Adding both contributions after 
appropriate rotation 
into the thin-slab basis [Fig.\ \ref{fig:directions}(c)] we obtain
\begin{multline}\label{eq:abthin}
    \frac{\bm{\Gamma}^{\textrm{ab,thin}}}{4\pi\eta\tau/\tilde{W}}  = \\  
  \hat{\bm{x}}\, \begin{pmatrix}
 \ln 2\,  \sin ^3\Delta\,  & 0 & 0 \\
 0 & \ln 2\,  \sin \Delta\,  \cos ^2\Delta\,  & i \frac{8}{3}   \chi \cos ^2\Delta\,  \\
 0 & -i \frac{8}{3}  \chi \cos ^2\Delta\,  & \ln 2\,  \sin \Delta\,  \\
   \end{pmatrix} \\ +\hat{\bm{y}}\,\sin\Delta 
  \begin{pmatrix}
 0 & \ln 2\,   \cos ^2\Delta\,  & -i \frac{8}{3} \chi \sin \Delta\,  \\
 \ln 2\,  \cos ^2\Delta\,  & 0 & 0 \\
 i \frac{8}{3}   \chi \sin \Delta\,  & 0 & 0 
  \end{pmatrix},
\end{multline}
where we
defined 
\[\Delta = \frac{\beta-\alpha}{2}.\]

To understand this result, it suffices to understand the 
current production due to arc-bulk excitations
at a single (bottom) surface, 
illustrated in Fig.\ \ref{fig:arc_bulk_transitions}.  
First we note that arc-bulk excitations vanish for 
the polarization component $x$ because
such a photon does not act on the spinor of the arc 
(which is an eigenspinor of $\sigma_x$) and thus 
cannot induce a transition to the orthogonal bulk state.
This is circumvented when the linear polarization points in the other
directions, $y$ and $z$. The induced velocity 
due to arc-bulk excitations sum up to a total velocity pointing 
in the $\hat{\bm{y}}$ direction, see 
Fig.\ \ref{fig:arc_bulk_transitions} (c), which explains the 
second term of \eqref{eq:resultab}. 

Circular $y + iz$ polarization instead acts like a ladder operator 
on the $\sigma_x$ eigenspinor and thus enhances the amplitude of
spin-flip excitations where the spin is increased (at 
positive $p_x$ in Fig.\ \ref{fig:arc_bulk_transitions}) and suppresses those where the spin is 
lowered (at negative $p_x$ in
Fig.\ \ref{fig:arc_bulk_transitions}), and vice versa for the 
opposite polarization handedness, $y-iz$, 
or chirality $\chi$ of the Weyl fermions. As is clear from Fig.\ 
\ref{fig:arc_bulk_transitions}, this asymmetry 
can produce  velocity in the
$x$ direction, which sign depends on the
polarization handedness
and the chirality. 
This explains the first term of  Eq.\ \eqref{eq:resultab}. 

\subsection{PGE due to bulk-bulk excitations}\label{sec:bulk_bulk}

In contrast to the arc-bulk excitations, the contribution of excitations within the 
bulk bands (from the valence bulk band to the 
conduction bulk band) strongly differ for the thin- and thick-slab cases. We thus 
consider the two cases separately.

\begin{figure}
    \centering
    \includegraphics[width=0.8\columnwidth]{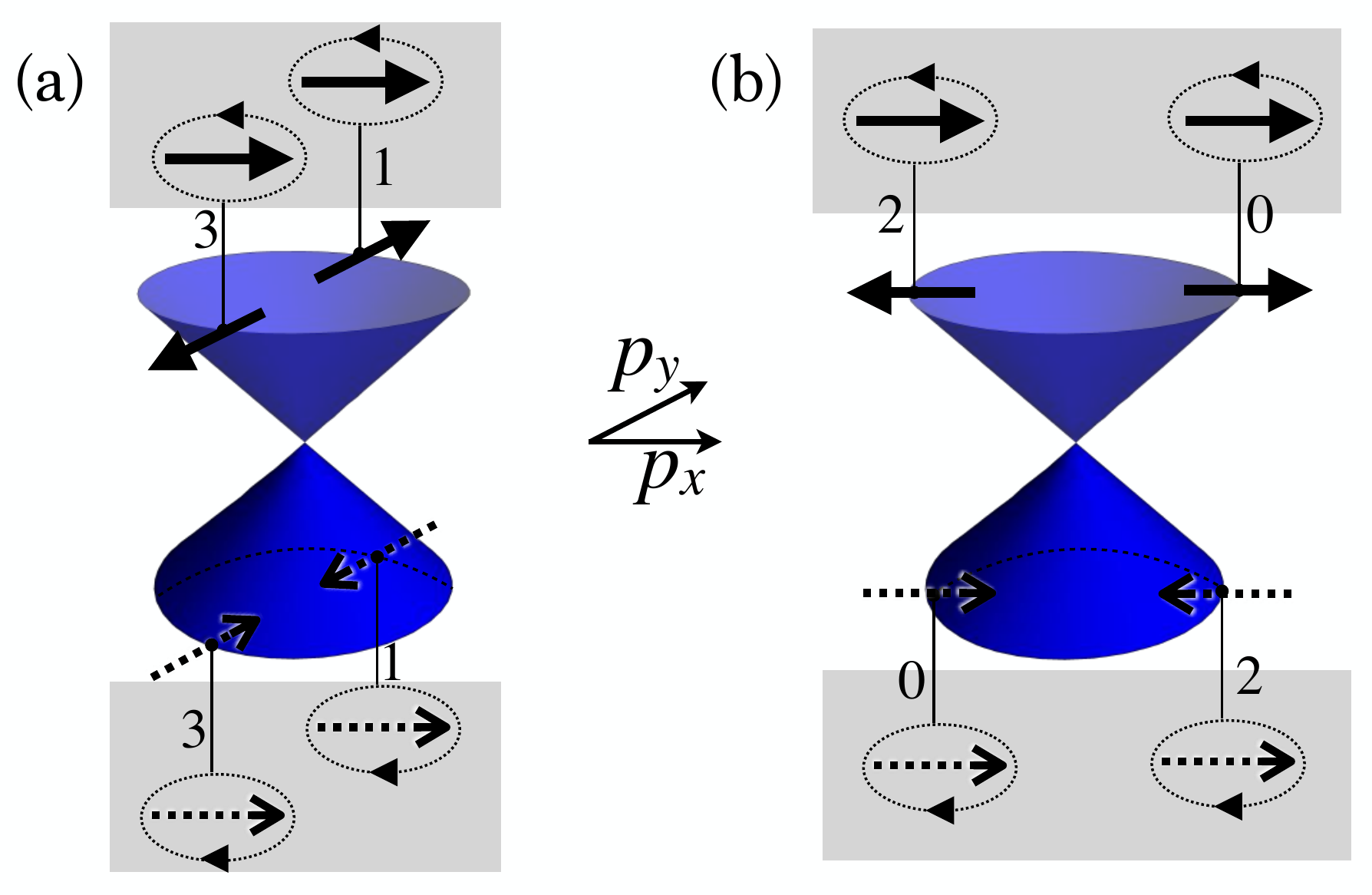}
    \caption{Characterization of bulk Weyl spinors  in a  
    semi-infinite spatial geometry. Spinors at the boundary are shown as arrows in the 
    gray areas. They point in the $x$ direction for all states. 
    Going away from the boundary
    the spinors rotate so that the average spin, indicated as arrows at the cones, 
    are like in an infinite system pointing parallel or antiparallel to the 
    momentum, depending on the band and chirality (here positive). The numbers
    characterize the angle between the boundary and the average spinor
    (in units of $\pi/2$), which differ by $\pi$ for opposite momenta.
    (a) and (b) show spinors at momenta perpendicular and parallel to the 
    boundary spinor, respectively.}
    \label{fig:bulk_bulk_transitions}
\end{figure}

\subsubsection{Thick slab limit $W \gg \delta$}

In this limit the light-induced excitations
are produced at a single (bottom) surface in 
the finite strip of width $\delta$. 
For the confinement-independent contribution 
$\bm{\Gamma}^{\textrm{bb}}$ we neglect all finite-size effects and obtain (see Appendix  \ref{app:calculation_bb_thick} 
for details)  
\begin{align}\label{eq:thickleading}
    \bm{\Gamma}^{\textrm{bb,thick}}_{ij} =&\ i\chi  \frac{2\pi \eta\tau }{3} \frac{\delta}{W}\ (\varepsilon_{xij}\hat{\bm{x}} + 
    \varepsilon_{yij}\hat{\bm{y}}).
\end{align}
Apart from  the absence of current in the direction perpendicular to the boundary and 
the factor $\delta/W$, this expression is identical to the circular PGE found in the infinite system model 
\cite{chan_photocurrents_2017}, which here has  been re-derived
using slab eigenstates. It is manifestly independent of the orientation of the Fermi arc. The prefactor 
$\delta/W$ correctly  reflects the fact that excitations 
occur in the fraction of the 
penetration depth of the full sample width.
 
The leading corrections 
in the thick-slab limit are of a higher order 
in $\ell/\delta$, see Table \ref{tab:length}. 
They stem from the $z$ integration 
in the matrix elements \eqref{eq:matelnew}, 
where we can still take the limit $W\to \infty$ but
keep the finite light penetration depth. Other finite-size 
corrections are controlled by the 
small parameter $\ell/W$ and can thus be neglected. Expansion 
to leading order in $\ell/\delta$
and numerical evaluation of the 
integral gives  (see Appendix  \ref{app:calculation_bb_thick} 
for details)  
\begin{multline}\label{eq:bbthick}
   \frac{ \delta \bm{\Gamma}^{\textrm{bb,thick}}}{\eta\tau/\tilde{W}}  \simeq    \hat{\bm{x}} \left(
\begin{array}{ccc}
 0 & 4.2 & 0 \\
 4.2 & 0 &  -16.8 i \chi \\
 0 &  16.8  i \chi& 0 \\
\end{array}
\right) \\
+ \hat{\bm{y}} 
  \left(
\begin{array}{ccc}
 -4.2 & 0 &  9.9 i \chi \\
 0 & -4.2 & 0 \\
 -9.9 i \chi & 0 & -8.4 \\
\end{array}
\right),
\end{multline}
written in the thick-slab basis, $\bm{\alpha}_1 =\hat{\bm{x}}$, 
$\bm{\alpha}_2 =\hat{\bm{y}}$. We estimate that these expressions are accurate to below $0.5\%$.
Corrections to the circular PGE (antisymmetric part of the tensor)
are in the same tensor components as the leading terms,
as they should according to the symmetry constraints. 
The corrections are of opposite sign as the leading 
contribution because the tendency of the boundary to 
align the initial and final spinor 
suppresses the transition amplitude 
(circular PGE needs spin-flip processes).  
A difference between $x$ and $y$ components
is a manifestation of the boundary-condition-broken symmetry between the $x$ and $y$
directions.

The response to linearly polarized 
light (symmetric part of the tensor) is something that is not found for an infinite-system Weyl cone in the absence of tilt. This follows from the fact that a single unconfined tiltless Weyl node is intra-node TR symmetric and hence there is no linear PGE. The vanishing of the linear PGE in such a system is due to cancellation of the  linearly-polarized-light-induced current from states at opposite momenta parallel to the polarization \cite{chan_photocurrents_2017}. While the 
symmetry considerations have already shown that linear
PGE contributions are possible in the presence of a boundary, (which breaks intra-node TR symmetry) it is peculiar that these contributions stem not only from arc-bulk but also 
from bulk-bulk excitations. To understand how the boundary
breaks the symmetry between
opposite momenta of bulk states, we consider the bulk-state 
spinor as a function of $z$, explicitly given in Eq.\ \eqref{eq:bwv}. 
At $z=0$
the boundary condition forces the spinors at all momenta to coincide with
$\psi(0) \propto (1,\exp[i\alpha])$. Going away from the 
boundary, the spinors rotate in the in-plane basis: 
At small $z$ the spinor can be written as $(1,\exp[ i\Phi(z)])$, 
with $\Phi(z)= -2 \arctan [(\pm\chi p-p_x)z]$. Since $p>|p_x|$, 
the rotation handedness is the same for all momenta and 
is set only by the chirality and the band ($\pm$). The spin averaged over 
the whole slab width coincides with the 
spin of an infinite system---parallel or antiparallel 
to the momentum, depending on the band and chirality. 
As illustrated in Fig.\ \ref{fig:bulk_bulk_transitions}, the angle between the
spinor at $z=0$ and the averaged spinor, measured in the direction of rotation,
thus always differs by $\pi$ for opposite momenta, which provides the 
crucial symmetry breaking and enables the response to linearly polarized radiation. 
Moreover, as can be seen from Eq.\ \eqref{eq:bwv} and
in Fig.\ \ref{fig:bulk_bulk_transitions}(b), the $x$ (i.e., $\bm{\alpha}_1$)
component of the spinor is invariant under simultaneous band change and 
$\bm{p}\to-\bm{p}$, which explains the vanishing diagonal components 
for the response in the $x$ direction.

\subsubsection{Thin slab $\delta \gg W \gg \ell$}

The confinement-independent contribution 
$\bm{\Gamma}^{\textrm{bb}}$ is obtained similarly to the
thick-slab case by neglecting all finite-size corrections. 
The only difference is that the integration 
over $z$ now extends over the whole slab width instead 
of $\delta$. The result,
\begin{align}\label{eq:thinleading}
    \bm{\Gamma}^{\textrm{bb,thin}}_{ij} =&\ i\chi  \frac{4\pi \eta\tau }{3}  \ (\varepsilon_{xij}\hat{\bm{x}} + 
    \varepsilon_{yij}\hat{\bm{y}}),
\end{align}
is, up to the missing factor 
$\delta/2W$, identical 
to the thick-slab case and, up to the 
vanishing current normal to the slab, identical 
to the known infinite-system result, as it should. 

For the confinement-induced contributions 
 we  collect finite-size corrections 
of the type $\ell/W$.
They stem from the quantization of 
$q_z$ and $p_z$ as well as from corrections to the wave functions and the velocity of bulk states.
We solve the problem numerically via discretizing the 
polar angle $\phi$, and finding $q_z, p_z$ pairs satisfying 
the energy conservation and Eq. \eqref{eq:bc_eq} using standard numerical tools (see Appendix \ref{app:calculation_bb_thin}  for details),  yielding 
\begin{widetext}
\begin{multline}\label{eq:response_tensor_thin}
     \frac{\delta\bm{\Gamma}^{\textrm{bb,thin}}}{\eta \tau/\tilde{W}} \simeq \bm{x} \left(
\begin{array}{ccc}
 -14.1 \sin \Delta + 4.7 \sin 3 \Delta & 0 & 0 \\
 0 & -21.5 \sin \Delta -4.7 \sin 3\Delta  & -i\chi(26.7+ 6.9 \cos 2\Delta ) \\
 0 & i\chi(26.7+ 6.9 \cos 2 \Delta ) & -23.0 \sin\Delta \\
\end{array}
\right)\\+\bm{y} \left(
\begin{array}{ccc}
 0 & 3.7 \sin \Delta -4.7 \sin 3\Delta & i\chi( 26.5 -  6.9 \cos 2\Delta ) \\
 3.7 \sin \Delta -4.7 \sin 3\Delta & 0 & 0 \\
 -i\chi( 26.5 -  6.9 \cos 2\Delta ) & 0 & 0 \\
\end{array}
\right),
\end{multline} 
\end{widetext}
where we defined
\begin{equation}
    \Delta \equiv \frac{\beta-\alpha}{2}.
\end{equation}
The numerical coefficients are accurate to the first decimal.
Together with Eq. \eqref{eq:resultab} and Eq. \eqref{eq:bbthick}, Eq. \eqref{eq:response_tensor_thin} represents the central quantitative result of this work. Due to  scale invariance of the Weyl Hamiltonian, these results are  generic for any Weyl semimetal with untilted Weyl cones,  up to straight-forward
directional rescaling in case of anisotropic velocity, as discussed in Appendix 
\ref{app:anisotropic_weyl_node}. 

\subsection{Discussion of confinement induced contributions and comparison to lattice simulation}
The arc-bulk contribution $\Gamma^{\textrm{ab}}$ and the confinement-induced bulk-bulk contribution $\delta\Gamma^{\textrm{bb}}$ are intimately linked: They are of the same order of magnitude and they always occur in combination. Therefore, only the sum $\delta\Gamma^{\textrm{bb}} + \Gamma^{\textrm{ab}}$ is experimentally relevant.

 In the thick-slab limit, the confinement-induced response tensor is
\begin{multline}\label{eq:confinement_pge_thick}
   \frac{ \delta \bm{\Gamma}^{\textrm{bb,thick}}+ \bm{\Gamma}^{\textrm{ab,thick}}}{\eta\tau/\tilde{W}}  \simeq \\
   \hat{\bm{x}} \left(
\begin{array}{ccc}
 0 & 4.2 & 0 \\
 4.2 & 0 & 0 \\
 0 & 0 & 0 \\
\end{array}
\right) 
+ \hat{\bm{y}} 
  \left(
\begin{array}{ccc}
 -4.2 & 0 &  9.9 i \chi \\
 0 & 0.2 & 0 \\
 -9.9 i \chi & 0 & -4.0 \\
\end{array}
\right).
\end{multline}
Note that since  $\Gamma^{\textrm{bb,thick}}_{xyz}$ and $\Gamma^{\textrm{ab,thick}}_{xyz}$ cancel (within numerical accuracy), circularly polarized light may produce a sizeable current only parallel to  the Fermi arc ($\bm{\alpha}_2=\hat{\bf{y}}$), whereas linearly polarized light may produce currents perpendicular to the Fermi arc as well. 

\begin{figure}
    \centering
    \includegraphics[width=\columnwidth]{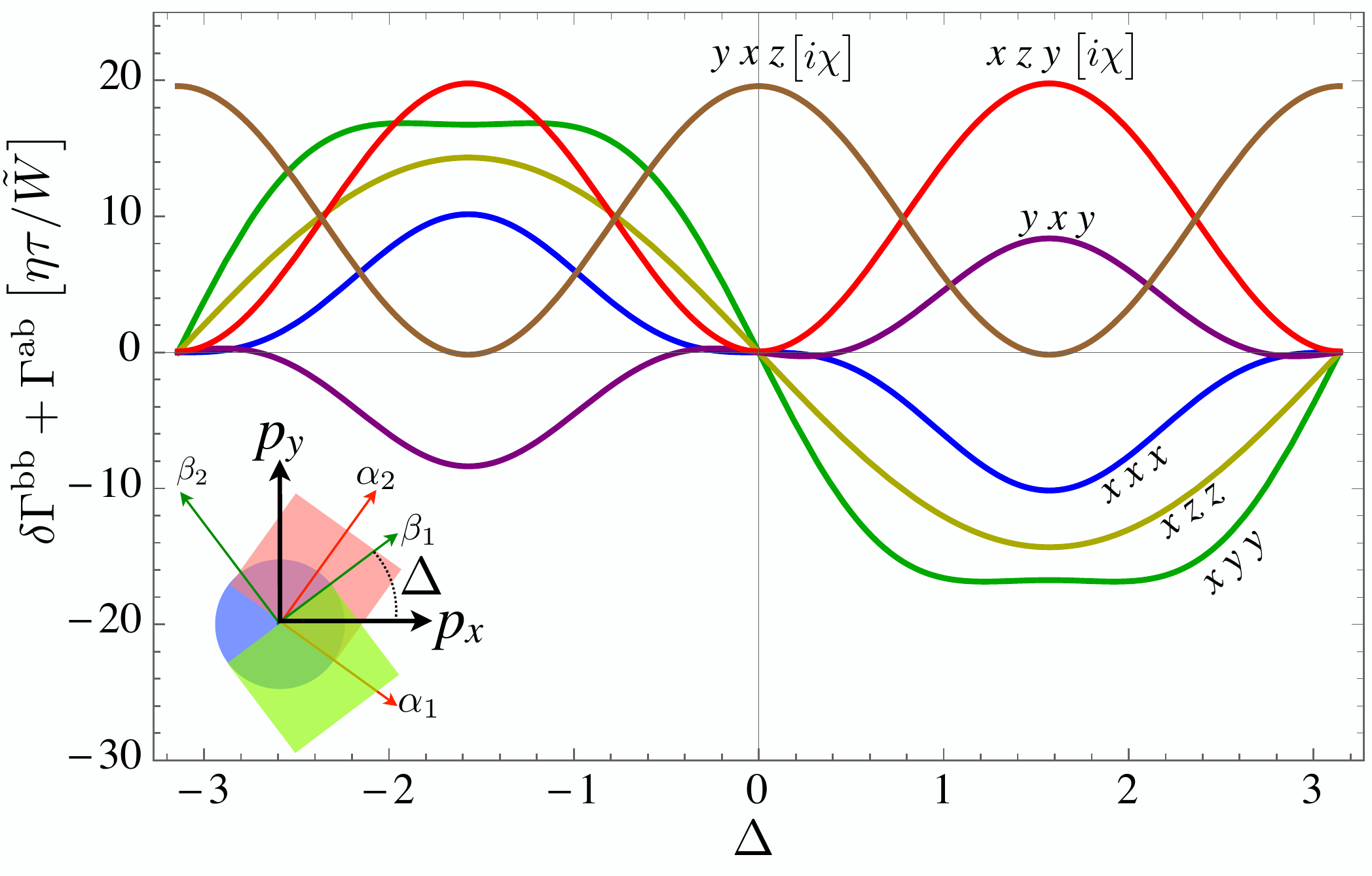}
    \caption{Confinement-induced response tensor $\delta\Gamma^{\textrm{bb,thin}}_{ijk}+\Gamma^{\textrm{ab,thin}}_{ijk}$ 
    as a function of $\Delta$. Only non-zero components
    are shown, labeled with $ijk$, the circular-PGE components
      have an additional prefactor  $i\chi$ as indicated. }
    \label{fig:thin_response_tensor}
\end{figure}

The thin slab limit result is plotted in Fig.\ \ref{fig:thin_response_tensor} as a function of $\Delta$.
For $\beta=\alpha$ ($\Delta = 0$), 
the linear PGE vanishes because
the second surface restores the symmetry between
opposite momenta: With regard to the 
corresponding discussion for the 
thick-slab limit, the sense of rotation of spinors 
away from the $z=W$ boundary is opposite to $z=0$ since 
$z$ runs ``backwards'' there. For $\beta = \pm\pi + \alpha$ ($\Delta = \pm \pi/2$) the symmetric part of the response tensor is approximately maximized, while the weight of circular response is simply shifted from one component to the other. The Fermi arc orientations thus change the nature of the response completely.

More generally, one can understand that the (anti)symmetric 
part of the tensor, i.e., the 
linear (circular) PGE, 
must be odd (even) in $\Delta$.
In terms of $\Delta$, the 
TR-breaking directions in 
\eqref{eq:Hhetero} are given by
$\bm{\alpha}_2 = (-\sin \Delta, \cos\Delta)$ and $\bm{\beta}_2 = (\sin \Delta, \cos\Delta)$. 
The transformation $\Delta \to -\Delta$ combined with the reflection $R_x$ and $\chi\to-\chi$ leaves the 
Hamiltonian invariant. From 
the corresponding transformation of the tensor follows that components of the symmetric part are odd while components of the anti-symmetric part are even
in $\Delta$, as seen in Fig.\ \ref{fig:thin_response_tensor}.

\begin{figure}
    \centering
    \includegraphics[width=\columnwidth]{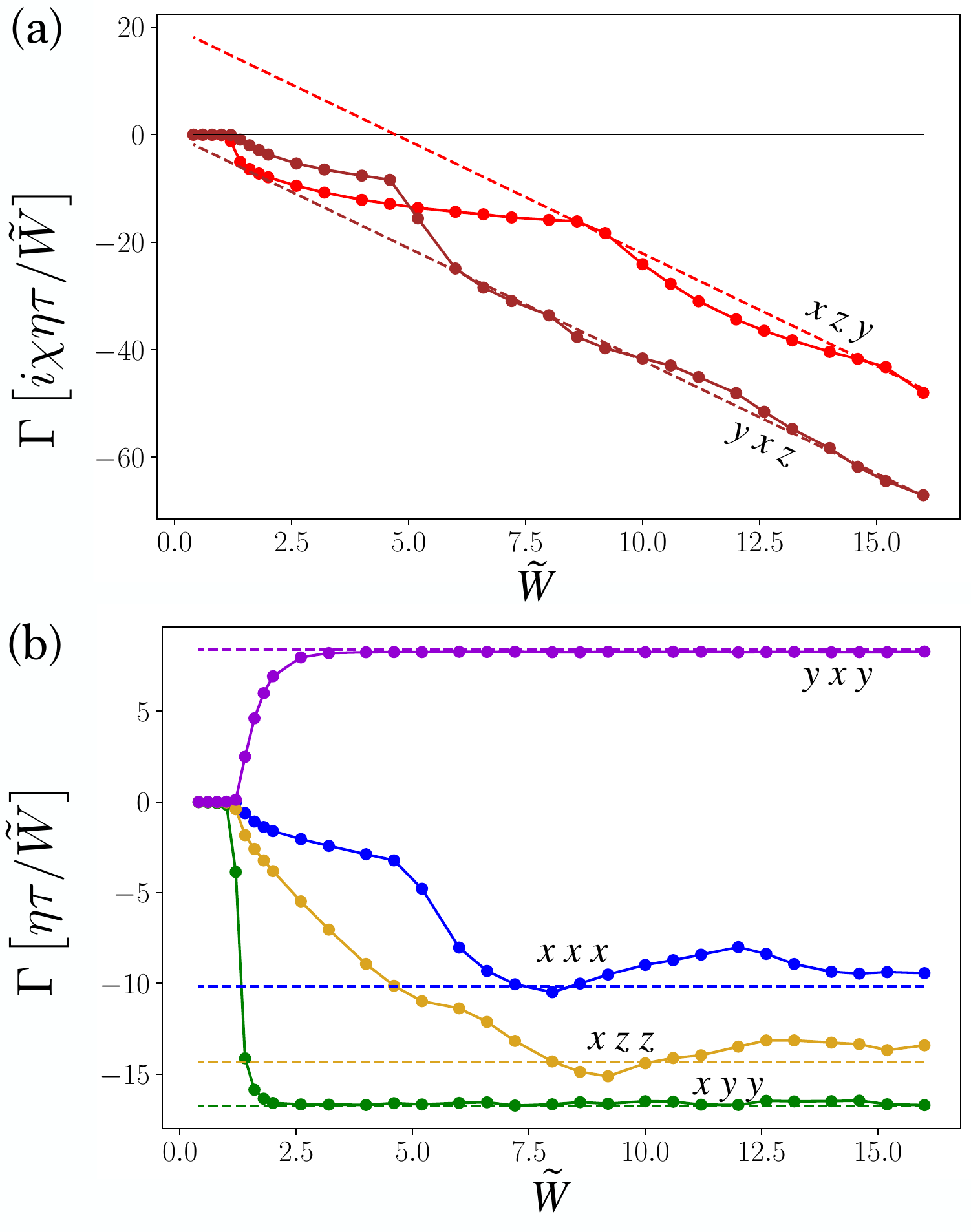}
     \caption{(a) Circular and (b) linear response tensor components in the ultra-thin limit $\tilde{W} = W/\ell \sim 1$ at $\Delta = \pi/2$. The data points correspond to the photogalvanic response tensor Eq. \eqref{eq:pge_formula_main} numerically evaluated for a lattice realization of a single Weyl point (see Appendix \ref{app:ultrathin} for details). The dashed lines correspond to the semi-analytical results $\Gamma = \Gamma^{\textrm{bb}} + \delta\Gamma^{\textrm{bb}} + \Gamma^{\textrm{ab}}$ in the limit $\tilde{W} \gg 1$. For $\tilde{W} \lesssim 2$, the lattice response vanishes as the frequency drops below the finite size gap.
    For $\tilde{W}\gtrsim 2$, the response converges 
    towards the semi-analytical $\tilde{W}\gg 1$ 
    results.
    \label{fig:lattice_results}}
\end{figure}

While for clarity of the analysis we considered the thin-slab case assuming $W\gg\ell$,  it is possible to 
relax this constraint and consider ultrathin slabs
with $W \sim \ell$ resorting to numerical techniques.
Here, we used a one-dimensional lattice realization of a single Weyl node (discretizing the $z$-direction while keeping $\bm{r}_{\parallel}$ continuous) to numerically evaluate the photogalvanic response tensor in Eq. \eqref{eq:pge_formula_main}. Details can be found in Appendix \ref{app:ultrathin}. The results are shown in Fig. \ref{fig:lattice_results}, demonstrating that our semi-analytical results can be reproduced in a lattice setting, and that the qualitative behaviour, such as sign and magnitude of the confinement induced contributions, extends down to $W\sim\ell$. 
For $W\lesssim 2\ell$, the finite-size gap of modes becomes larger than the photon energy and the response vanishes.

 \subsection{Centrosymmetric Weyl semimetal}
\label{sec:WSM}

\begin{figure}
    \centering
    \includegraphics[width=\columnwidth]{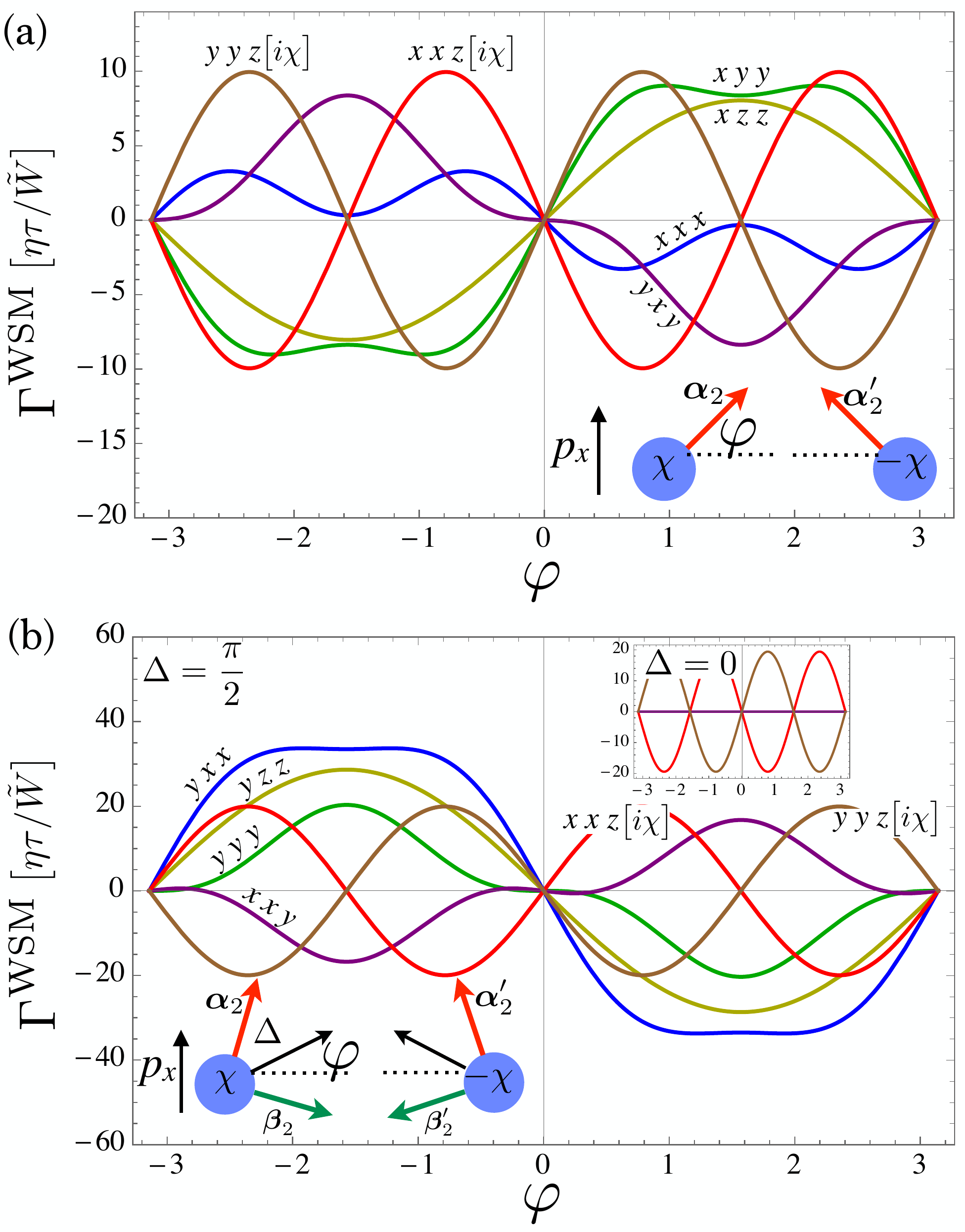}
    \caption{Response-tensor components
    for a centrosymmetric WSM in 
    (a) the
    thick- and (b) the thin-slab case,
    as a function the angle  $\varphi$. As shown in the lower insets, $\phi$ sets the deviation from 
    antiparallel alignment of (a) the 
    direction of bottom-surface Fermi arcs; and (b) the directions of bisectors
    of bottom- and top-surface Fermi arcs. In the
    thin-slab case (b) the angle between 
    top and bottom arcs is chosen as 
    $\Delta=\pi/2$; the inset in the
    right top shows the result for
    $\Delta=0$, for which the linear PGE vanishes.}
    \label{fig:WSM}
\end{figure}

Our results for a single Weyl node allow to infer on the 
response of a WSM with several nodes by adding the 
contributions of each node.
Probably the most intersting case is that of 
centrosymmetric WSMs, for which the confinement-independent bulk-bulk 
contributions $\Gamma^{\textrm{bb}}$ cancel and only the confinement-induced contributions, $\delta \Gamma^{\textrm{bb}}$ and $\Gamma^{\textrm{ab}}$, survive. 
A minimal model of a centrosymmetric bulk WSM consists of 
a single pair of Weyl nodes with opposite chirality. 
Considering the multilayer Hamiltonian 
$H_{\chi m}$  in \eqref{eq:Hhetero}
as the Hamiltonian describing one of the Weyl nodes, for the 
Hamiltonian describing the second Weyl node of opposite chirality we take 
$H_{ -\chi -m}$. In this case the Fermi arcs emanate in 
opposite directions, which happens when the Weyl nodes 
are connected in 
a straight line. However, 
we have seen in Sec.\ \ref{sec:model}
that an additional boundary potential 
 $\delta H_\mathrm{b} = 
\delta(z) \mu_0 + \delta(z-W)\mu_W$ rotates the spinors 
by $\alpha\to\alpha+\chi 2\mu_0$ and 
$\beta\to\beta-\chi 2\mu_W$. The generic situation is
thus that there is a finite angle $\varphi$ 
characterizing the deviation from an anti-parallel 
alignment, as shown in the inset of 
Fig.\ \ref{fig:WSM}(a). 
Note that between the nodes, the Fermi arcs thus must be curved,
as is typically seen in experiments; the curvature itself
plays however no role for our results since 
excitations occur only close to the Weyl nodes. 

The total response
$ \Gamma^\mathrm{WSM}_{abc}$  is obtained 
from the single-cone result 
$\Gamma_{ijk}(\chi,m)$ (now explicitly 
denoting the $\chi,m$ dependence),
\begin{multline}
 \Gamma^\mathrm{WSM}_{abc}=   R(\varphi)_{ai} R(\varphi)_{bj} R(\varphi)_{ck}\Gamma_{ijk}(\chi, m)\\
    +R(-\varphi)_{ai} R(-\varphi)_{bj} R(-\varphi)_{ck}\Gamma_{ijk}(-\chi,-m),
\end{multline}
where $R(\varphi)$ is the spatial rotation matrix
for a rotation around $z$ by $\varphi$. The results are plotted in 
Fig.\ \ref{fig:WSM}. 
From 
the transformation 
behavior of the response tensor discussed in Sec.\ \ref{sec:symmetry}
(symmetric part odd in $m$ and 
even in $\chi$, antisymmetric part odd in $\chi$ and even in $m$), the response of the two Weyl nodes cancel each other at $\phi=0$. This means that
in the thick-slab case the response vanishes when the
Fermi arcs of the illuminated surface emanate from the 
Weyl nodes in exactly opposite directions. In the thin-slab case, the same applies but the emanation
direction is replaced by the bisector of the top- and bottom-surface Fermi arcs.

For $\phi=\pi/2$ the directions
 just discussed (emanation direction for thick slab 
 and bisector direction for thin slab) are  parallel.
 This is equivalent to taking the contributions 
 of the two Weyl nodes at the same 
 $m$ (instead of $m$ and $-m$), while 
 $\chi$ are still opposite. 
 Since the (anti)symmetric 
part is even (odd) in $\chi$, the 
antisymmetric parts cancel also here
but the symmetric parts add up to 
twice the value of a single cone.
This can 
be seen by comparison of Fig.\ 
\ref{fig:WSM}(b) with the single-cone results shown in Fig.\ \ref{fig:thin_response_tensor} for the thin-slab 
case and Fig.\ 
\ref{fig:WSM}(a) with 
$\bm{\Gamma}^{\textrm{ab,thick}}+
\delta\bm{\Gamma}^{\textrm{bb,thick}}$ from Eqs.\ \eqref{eq:resultab} and 
\eqref{eq:bbthick} for the thick-slab
case.
(Note that 
the coordinate system is now rotated by 
$\pi/2$, i.e., $y\to x$ and $x\to-y$, compared to the single-cone case).

\section{Conclusion}\label{sec:conclusion}

In conclusion, we have explored the PGE of 
a WSM spatially confined to a 
slab geometry. Symmetry breaking 
on the surfaces via the orientation of the Fermi arcs enables
circular and  
linear photogalvanic response currents, which would 
otherwise not be 
possible, in particular, in centrosymmetric WSMs.

The magnitude of the confinement-enabled PGE  inherits the topology-enhancement of an 
unconfined Weyl Fermion, based on the topologically protected band touchings. 
However, while in infinite systems those band 
touchings include only the chiral pairs of 
Weyl nodes, a confined system features 
topological surface states, which are 
tightly glued to the Weyl nodes.

The ratio of confinement-induced 
contributions to bulk
contributions scales in case of 
a thin slab like  $(v/c)\times(\lambda/W)$
and for the thick slab 
like 
$v/c$, where $\lambda$ is the light wavelength, $W$ the slab thickness, and
$v/c\approx 0.01$ the node- vs.\ light-velocity. Considering the upper and lower bounds of 
$\lambda$ set by the finite Fermi level and 
the band width of 
typical WSM materials, the
confinement-induced PGE is 
on the order of bulk PGE
for widths of order
$W\sim 0.1\dots 1\, \mu$m. Surface-controlled PGE  is thus found in such thin slabs even for non-centrosymmetric WSMs, which makes the experimental realization of thin WSM slabs
or even stacks of those especially interesting. 

One of the most remarkable properties
of the confinement-induced PGE is that it is
controlled by surface boundary conditions. We explicitly discussed the effect of a surface potential, which rotates the direction 
of Fermi arcs.
Another interesting 
possibility, known from experiments,
are layered WSMs for which the different surface terminations can
not only change the directionality of 
Fermi arcs but even lead 
to different connectivities to the 
Weyl nodes 
\cite{Morali2019,Fujii2021}. This Fermi-arc geometry is 
observable, e.g., via angle-resolved photoemission spectroscopy \cite{Xu2015b, Xu2015,yang_weyl_2015};  our work links this
geometry with the photogalvanic response. 

With regard to the remarkable recent
progress in device 
microstructuring \cite{Moll2018, Zhang2019, Nishihaya2019, Bedoya-Pinto2020,Qin2020}, our 
work might thus play an important role in
identifying Weyl physics and shaping 
the photogalvanic response by designing 
the material surface.

\subsubsection*{Note added}
When this work was substantially completed, we became aware of a recent article \cite{Cao2021}
considering the PGE of a finite system using a minimal centrosymmetric lattice model
that features two Weyl nodes. 
This article focuses on Fermi energies
far away from the Weyl nodes where 
the PGE is governed by non-linear terms of the
dispersion, while in the semimetallic 
regime, which is the focus of our work, 
the response of their model vanishes as it, 
in terms of our model, 
considers only the specific $\varphi=0$ case
of Sec.\ \ref{sec:WSM}. 
This trivial case is the only place of overlap of our works.

\acknowledgements
We benefited from a discussion with Haim Beidenkopf. 
J.~F.~S.\ gratefully acknowledges financial support by QuantERA project {\em Topoquant},
as well as by the Deutsche Forschungsgemeinschaft (DFG, German Research Foundation) through Priority Program 1666 and CRC 910. 
M.~B.\ was funded by Grant No. 18688556 of the DFG. The work of A.~A.\ was supported by the National Science Foundation
Grant MRSEC DMR-1719797.

\bibliography{refs}

\onecolumngrid

\appendix

\section{Confined Weyl Fermion wave functions}\label{app:wave-functions}
In this appendix we derive the explicit form of the eigenfunctions of Eq. \eqref{eq:ham_0} in the slab geometry  $0<z<W$. Writing the momentum operator $p_z$ in the spatial basis, $p_z= -i\partial_z$ the 
Weyl equation $H\psi(z)=E\psi(z)$ 
can be written as
\begin{align}
    -i\partial_z \psi(z) = \mathcal{P}_z \psi(z),
\end{align}
where the generalized momentum $\mathcal{P}_z$ was defined in Eq. \eqref{eq:gen_mom}. 
The Weyl equation is formally solved by Eq. \eqref{eq:wv0}. 
Defining an orthonormal basis for our model of  zero out-of-plane current states,
\begin{align}
    \ket{\alpha_\pm} =&\ \frac{1}{\sqrt{2}}\begin{pmatrix} 1 \\ \pm e^{i\alpha} \end{pmatrix},\ 
    \ket{\beta_\pm} = \frac{1}{\sqrt{2}}\begin{pmatrix} 1 \\ \pm e^{i\beta} \end{pmatrix},
\end{align}
the generic boundary conditions at the two surfaces can be written as
\begin{equation}\label{eq:bc_app}
    \psi(0) \propto\ket{\alpha_+}, \;\;\;\;\;
    \psi(W) \propto \ket{\beta_+}.
\end{equation}
From the boundary conditions, the quantized 
eigenvalues $p_z$ are the solutions of 
\begin{align} \label{eq:bc_eq_app}
 0  = \sin \frac{\alpha-\beta}{2} + \frac{\tan (p_z W)}{p_z} \Bigg[ & p_x \mp \chi p \cos \frac{\alpha-\beta}{2} \Bigg],
\end{align}
where $p_x$ is in the basis of 
Fig.\ \ref{fig:directions}(c).
Solutions with real $p_z$ correspond to bulk states, imaginary solutions correspond to surface ``arc" states, which we 
will now discuss in more detail.

\subsection{Arc states}
For an imaginary $p_z$, normalizable wavefunctions
are found that are localized at the bottom (b) and top (t) surfaces, 
\begin{subequations}\label{eq:arc_wf}
\begin{align}
    \psi_{\textrm{arc}}^{\mathrm{b}}(z) =&\ \sqrt{2 \bm{p}\cdot\bm{\alpha}_2}\ 
    e^{-  \bm{p}\cdot\bm{\alpha}_2 z } \ket{\alpha_+} = \braket{z}{\textrm{arc,b},\bm{p}_{\parallel}}, \label{eq:arc_b}\\
    \psi_{\textrm{arc}}^{\mathrm{t}}(z) =&\ \sqrt{-2 \bm{p}\cdot\bm{\beta}_2 }\ e^{  \bm{p}\cdot\bm{\beta}_2 (W-z) }  \ket{\beta_+} =  \braket{z}{\textrm{arc,t},\bm{p}_{\parallel}}.
    \label{eq:arc_t}
\end{align}
\end{subequations}
(These expressions assume $W\abs{\bm{x}_2 \cdot \bm{p}} \gg 1$ for $x\in\{\alpha,\beta\}$).
From the criterion of normalizability, the momenta of Fermi arcs are bound to 
\begin{align}
      \bm{p}\cdot\bm{\alpha}_2  >0 ,\ \ \ \  \bm{p}\cdot\bm{\beta}_2 < 0.
\end{align}
The dispersion relations read 
\begin{align}\label{eq:arcdisp}
    E^{\mathrm{b}}_{\textrm{arc}} =&\ \chi v \bm{p}\cdot\bm{\alpha}_1,\  
    E^{\mathrm{t}}_{\textrm{arc}} = \chi v \bm{p}\cdot\bm{\beta}_1,
\end{align}
and hence the velocity expectation values are
\begin{align}
    \bm{v}^{\mathrm{b}}_{\textrm{arc}} =&\ \chi v 
    \bm{\alpha}_1,\ 
    \bm{v}^{\mathrm{t}}_{\textrm{arc}} = \chi v \bm{\beta}_1.
\end{align}
The directions $\bm{\alpha}_i$ and
$\bm{\beta}_i$
are thus the directions in which the Fermi arcs emanate from the Weyl node
($i=1$) and the directions of their motion ($i=2$), both up to 
the sign, as indicated in Fig.\ \ref{fig:directions}. 

\subsection{Bulk states}
For a real $p_z$, from 
\eqref{eq:wv0} the normalized wavefunctions 
of the conduction $(+)$ and valence band $(-)$ read
\begin{align}
    \psi_{\bm{p}\pm}(z) =&  \frac{1}{\sqrt{W N_{\bm{p}\pm}}}
     \Big\{ \bqty{p_z \cos p_z z - \bm{\alpha}_2\cdot\bm{p} \sin p_z z}\ket{\alpha_+} 
      + i\bqty{\pm\chi p- \bm{\alpha}_1\cdot\bm{p} } \sin p_z z \ket{\alpha_-} \Big\} =  \braket{z}{\textrm{bulk},\pm,\bm{p}} ,
      \label{eq:bwv}
\end{align}
where the normalization factor, 
isolating the finite-size correction $\sim 1/W$, 
is given by 
\begin{subequations} \label{eq:normalization}
\begin{align}
     N_{\bm{p}\pm} =&\ p( p \mp \chi \bm{\alpha}_1\cdot\bm{p}) + \delta N_{\bm{p}\pm},\\
     \delta N_{\bm{p}\pm} =&\ -\frac{\sin (W p_z) }{W p_z} \Big[ \pqty{p_{\parallel}^2 \mp \chi p\ \bm{\alpha}_1\cdot\bm{p}  } \cos  (W p_z) 
      +  p_z \bm{\alpha}_2\cdot\bm{p} \sin  (W p_z) \Big]. 
\end{align}
\end{subequations}
The velocity expectation values read
\begin{subequations}\label{eq:velocity_general}
\begin{align}
     \bm{v}_{\bm{p}\pm}  =&\  \pm v  \frac{\bm{p}_{\parallel}}{p} +  \delta \bm{v}_{\bm{p}\pm}, \\ 
    \delta \bm{v}_{\bm{p}\pm} =&\  \frac{\partial E_{\bm{p}\pm}}{\partial p_z} \frac{\textrm{d} p_z}{\textrm{d} \bm{p}}  =\pm v   \frac{p_z}{p} \frac{\textrm{d} p_z}{\textrm{d} \bm{p}},
\end{align}
\end{subequations}
where we again isolated the finite-size correction 
$ \delta \bm{v}_{\bm{p}\pm}\sim 1/W$, which stems from the weak 
$\bm{p}_\parallel$ dependence of 
the quantized $p_z$, as implicitly 
given in \eqref{eq:bc_eq}. Note that $v_z = 0$ due to the boundary conditions for all states.

\section{Anisotropic Weyl node}
\label{app:anisotropic_weyl_node}

Here we generalize our calculations 
to Weyl fermions with anisotropic 
velocity. We consider the Hamiltonian
\begin{equation}
    h'(\bm{p}) = v_{ij} \sigma_i p_j \equiv \chi v \bm{p}' \cdot \bm{\sigma} = h(\bm{p}'),
\end{equation} 
where $v_{ij} = v_{ji}$ and we defined $p'_i = \tilde{v}_{ij} p_j$ in terms of 
\begin{equation}
    \tilde{v}_{ij} = \chi \frac{v_{ij}}{v},\ v > 0.
\end{equation}
The chirality of the anisotropic Weyl node is $\chi =  \textrm{sign} ( \det v)$. Undashed symbols refer to the isotropic case discussed in the main text. The current operator is 
\begin{equation}
    j'_i = - e v_{ij} \sigma_j = \tilde{v}_{ij} j_j.
\end{equation}
To avoid complications in the boundary condition we specify to
\begin{equation}
     v = \textrm{diag}(v_{\parallel} , v_z),
\end{equation}
where $v_{\parallel}$ is a 2x2 matrix acting only on components parallel to the boundary. Furthermore, we set $v = \abs{v_z}$. With this we may again employ the boundary conditions of Eq. \eqref{eq:bc}. Then, the arc and bulk wave-functions may be obtained by simply replacing $\bm{p}  \to \bm{p}'$ in Eqs. \eqref{eq:arc_b}, \eqref{eq:arc_t} and Eq. \eqref{eq:bwv}. Similarly, the velocities can be expressed in terms of the isotropic expressions via 
\begin{equation}
    \pqty{\bm{v}'_{\bm{p},n}}_i = \frac{\textrm{d} E'_{\bm{p},n}}{ \textrm{d} p_i } = \frac{\partial p'_i}{\partial p_i} \frac{\textrm{d} E_{\bm{p}',n}} {\textrm{d} p'_i } = \tilde{v}_{ij}\pqty{\bm{v}_{\bm{p}',n}}_j.
\end{equation}
Altogether, the response tensor of the anisotropic Weyl node is related to the isotropic node via (c.f. Eq. \eqref{eq:pge_formula_main})
\begin{subequations}
\begin{align}
\Gamma'_{ijk} =&\  \frac{8\pi\eta \tau}{  W} 
    \int d^2 p_{\parallel} \, \sum_{q_z p_z }  
    \pqty{\frac{\bm{v}'_{\bm{p}+}}{v}  
     -\frac{ \bm{v}'_{\bm{q}-}}{v} }_i     \delta 
\pqty{1 - \frac{E'_{\bm{p}}}{\omega }
-\frac{E'_{\bm{q}}}{\omega}}
\bqty{\bm{M}'_{\bm{p}\bm{q}} \otimes (\bm{M}'_{\bm{p}\bm{q}})^*}_{jk} \\
=&\ \frac{\tilde{v}_{il}\tilde{v}_{jm}\tilde{v}_{kn}}{\abs{\det \tilde{v}}}  \frac{8\pi\eta \tau}{ W } 
    \int d^2 p'_{\parallel} \, \sum_{q'_z p'_z }  
    \pqty{\frac{\bm{v}_{\bm{p}'+}}{v}  
     -\frac{ \bm{v}_{\bm{q}'-}}{v} }_l     \delta 
\pqty{1 - \frac{E_{\bm{p}'}}{ \omega }
-\frac{E_{\bm{q}'}}{ \omega}}
\bqty{\bm{M}_{\bm{p}'\bm{q}'} \otimes  \bm{M}_{\bm{p}'\bm{q}'}^*}_{mn} \\ 
=&\  \frac{\tilde{v}_{il}\tilde{v}_{jm}\tilde{v}_{kn}}{\abs{\det \tilde{v}}}    \Gamma_{lmn},
\label{eq:pge_formula_aniso}
\end{align}
\end{subequations}
where the determinant stems from the change of variables $\bm{p}_{\parallel}$ to $\bm{p}_{\parallel}'$ (using $\abs{\det \tilde{v}_{\parallel}} = \abs{ \det \tilde{v}}$ due to our choice of $v$).

\section{Symmetry 
considerations}\label{app:symmetry}

\subsection{Thick
slab}

We work in the 
basis of the thick slab $\bm{\alpha}_1 = 
\hat{\bm{x}},\ \bm{\alpha}_2 = 
\hat{\bm{y}}$. 
The heterostructure Hamiltonian Eq.
\eqref{eq:Hhetero} of the semi-infinite system under consideration 
can be written as
\begin{equation}
    H_{\chi m} =  \chi v \bm{p}
\cdot \bm{\sigma} +\begin{cases}
- \chi m \sigma_y & z<0 \\ 0 & z>0
\end{cases}.
\end{equation}
where the vacuum at $z<0$ is modeled by a large mass 
term  with $m\to\infty$, which acts like a magnetization in breaking 
the intra-node TR symmetry, as discussed in the main text.
We consider spatial mirror-plane 
reflections $R_i$, $i=x,y,z$, with 
$i=x$ corresponding to reflection w.r.t.\
the $yz$ plane, etc. A single reflection 
 reverses the component of the momentum
and the current that is normal to the mirror plane and the components  of the spin that are parallel to the mirror plane.
The action of the reflections  on the Hamiltonian thus read
\begin{equation}
    R_xH_{\chi m}R_x = H_{-\chi m},\ \ \ \ \ \  \ \ \ 
    R_yH_{\chi m}R_y = H_{-\chi -m}.
\end{equation}
In words, reflection $R_x$ reverses the chirality and reflection 
$R_y$ reverses the chirality and the magnetization.

From this we can infer on the transformation behavior 
of the response tensor.  First, note that the (anti)symmetric part 
of the response tensor,  $\Gamma^\mathrm{s}$
($\Gamma^\mathrm{a}$),
is generally odd (even) under TR
\cite{belinicher_photogalvanic_1980}
-- the (anti)symmetric  part is thus odd (even) under $m\to -m$. Second, in Sec.\ 
\ref{sec:symmetry} we have shown that the (anti)symmetric part 
of the response tensor is even (odd) under $\chi\to-\chi$. Taking also into account 
the transformation of the current under reflections, we 
 obtain for the symmetric part,
 \begin{subequations}
\begin{align}
    R_x\Gamma^\mathrm{s}_xR_x =&\ -\Gamma^\mathrm{s}_x, \\ 
    R_x\Gamma^\mathrm{s}_y R_x =&\ \Gamma^\mathrm{s}_y, \\ 
    R_y\Gamma^\mathrm{s}_xR_y =&\ -\Gamma^\mathrm{s}_x, \\ 
    R_y\Gamma^\mathrm{s}_yR_y =&\ \Gamma^\mathrm{s}_y, 
\end{align}
\end{subequations}
while the antisymmetric part satisfies
\begin{subequations}
\begin{align}
    R_x\Gamma^\mathrm{as}_x R_x =&\ \Gamma^\mathrm{as}_x, \\ 
    R_x\Gamma^\mathrm{as}_y R_x =&\ - \Gamma^\mathrm{as}_y, \\ 
    R_y\Gamma^\mathrm{as}_xR_y =&\ - \Gamma^\mathrm{as}_x, \\ 
    R_y\Gamma^\mathrm{as}_yR_y =&\ \Gamma^\mathrm{as}_y. 
\end{align}
\end{subequations}
From this follows 
\begin{align}
\Gamma^{\textrm{thick}}_x  
=&\ \begin{pmatrix}
0& \Gamma_{xxy}  & 0 \\
\Gamma_{xxy} & 0&  \Gamma_{xyz} \\
0& -\Gamma_{xyz} & 0 \end{pmatrix},\ 
\Gamma^{\textrm{thick}}_y  
= \begin{pmatrix}
\Gamma_{yxx} & 0  & - \Gamma_{yzx} \\
0 & \Gamma_{yyy} & 0 \\
\Gamma_{yzx}& 0 & \Gamma_{yzz} \end{pmatrix}.
\end{align}

\subsection{Thin slab}

We now work in the 
basis of the thin slab, depiceted in 
Fig.\ \ref{fig:directions}(c). We now consider the full heterostructure Hamiltonian
\eqref{eq:Hhetero}, 
\begin{equation}
    H_{\chi m } =  \chi v \bm{p}
\cdot \bm{\sigma} +\begin{cases}
- \chi m\, \bm{\sigma}\cdot\bm{\alpha}_2 & z<0 \\ 0 & 0<z<W \\
 \chi m\, \bm{\sigma}\cdot\bm{\beta}_2 & z>W
\end{cases},
\end{equation}
with $m \to\infty$.  The behavior under reflections and  TR is as in the previous section but now with two TR-breaking 
magnetizations. 

We consider two symmetry transformations. First we note that the 
combined reflection $R_z$, which 
swaps the top and bottom surfaces, and $R_y$, which interchanges $\bm{\sigma}\cdot\bm{\alpha}_2$ and $\bm{\sigma}\cdot\bm{\beta}_2$, leave the
Hamiltonian invariant. Second, the combination of $R_x$ and $R_y$ inverts both 
$\bm{\sigma}\cdot\bm{\alpha}_2$ and $\bm{\sigma}\cdot\bm{\beta}_2$,
which can be compensated by $m\to-m$. Altogether, 
\begin{equation}
    R_yR_z  H_{\chi m } R_zR_y =  H_{\chi m },\ \ \ \ \ \ \ \ 
    R_xR_y H_{\chi m } R_y R_x =  H_{\chi -m}.
\end{equation}

Both the symmetric and 
the antisymmetric parts of the response tensor thus satisfy
\begin{subequations}
\begin{align}
    R_yR_z \Gamma_x R_zR_y =&\ \Gamma_x, \\ 
    R_yR_z \Gamma_y R_zR_y =&\ -\Gamma_y,
\end{align}
\end{subequations}
and, since the symmetric part is odd under $m\to-m$, it satisfies
\begin{subequations}
\begin{align}
    R_xR_y \Gamma^\mathrm{s}_x R_yR_x =&\ \Gamma^\mathrm{s}_x, \\ 
    R_xR_y \Gamma^\mathrm{s}_y R_yR_x =&\ \Gamma^\mathrm{s}_y,
\end{align}
\end{subequations}
while the antisymmetric part satisfies
\begin{subequations}
\begin{align}
    R_xR_y \Gamma^\mathrm{as}_x R_xR_y =&\ -\Gamma^\mathrm{as}_x, \\ 
    R_xR_y \Gamma^\mathrm{as}_y R_xR_y =&\ -\Gamma^\mathrm{as}_y.
\end{align}
\end{subequations}
From this follows,
\begin{align}
\Gamma^{\textrm{thin}}_x  
=&\ \begin{pmatrix}
\Gamma_{xxx} & 0  & 0 \\
0&  \Gamma_{xyy} &  \Gamma_{xyz} \\
0& -\Gamma_{xyz} & \Gamma_{xzz} \end{pmatrix},\ 
\Gamma^{\textrm{thin}}_y  
= \begin{pmatrix}
0 & \Gamma_{yxy}  & - \Gamma_{yzx} \\
 \Gamma_{yxy} & 0 & 0 \\
\Gamma_{yzx}& 0 & 0 \end{pmatrix}.
\end{align}

Finally, we may also constrain the dependency of the components with respect to the angle $\Delta = (\beta - \alpha)/2 \in [-\pi,\pi]$. In terms of $\Delta$, the magnetization directions are given by
$\bm{\alpha}_2 = (-\sin \Delta, \cos\Delta)$ and $\bm{\beta}_2 = (\sin \Delta, \cos\Delta)$. First, note that the transformation $\Delta \to -\Delta$ may be compensated by $R_x$. From this follows that components of the symmetric part of $\Gamma$ are odd under $\Delta \to -\Delta$ while components of the anti-symmetric part are even. 
Next note, that $\Delta \to \Delta + \pi$ inverts both magnetizations and may be compensated by TR. From this follows that components of the symmetric part of $\Gamma$ are odd under $\Delta \to \Delta + \pi$ while components of the anti-symmetric part are even.

\section{Photogalvanic current due to arc-bulk excitations}\label{app:calculation_ab}
As explained in the main text, the leading-order current
contribution due to arc-bulk excitations is of the same order of magnitude as the subleading contributions from bulk-bulk excitations. 
Other types of finite-size corrections that we had to account for when considering
bulk-bulk excitations can now be disregarded, as they would give corrections 
of higher order. In particular, we 
can replace the momentum sum over $p_z$ by integrals in both the thick and thin slab regimes. Furthermore, we can assume $W \to \infty$ and $\delta \to \infty$ in both regimes and neglect the finite light momentum $\bm{k}$. In the thick slab limit only the bottom arc is relevant, in the thin slab limit both arcs contribute.

The response tensor due to the bottom arc is
\begin{align}\label{eq:ab_current_bot_app}
    \bm{\Gamma}^{\textrm{ab,b}}_{ij} = \frac{8 \eta \tau}{W} 
    \int \frac{ d^3p}{p} \Theta( p_z ) \Theta( p_y ) 
    \Bigg\{&
    \Theta(-\chi p_x) \delta\pqty{1 - p + \chi p_x } 
    \bqty{ \frac{\bm{p}_{\parallel}}{p} - \chi \bm{\alpha}_1 }
    M^{\textrm{ab,b},+}_i \pqty{\bm{M}^{\textrm{ab,b},+}_j}^* \nonumber \\ 
    &+ 
    \Theta(\chi p_x) \delta\pqty{1 - \chi p_x - p} 
    \bqty{ \chi \bm{\alpha}_1 + \frac{\bm{p}_{\parallel}}{p} }
    M^{\textrm{ab,b},-}_i \pqty{\bm{M}^{\textrm{ab,b},-}_j}^* 
     \Bigg\}
\end{align}
where all momenta are dimensionless 
(in the appendix we suppress the tilde, which denotes dimensionless units in the main text). The first line captures transitions between conduction band and arc sheet, while the second captures transitions between the arc sheet and the valence band (the first Heaviside-$\Theta$ function enforces normalizability of the arc states, the second allows transitions between empty and occupied states only). We also defined the arc-bulk matrix elements ($\bm{p} = p (\sin\theta\cos\phi,\sin\theta\sin\phi,\cos\theta)$)
\begin{subequations}
\begin{align}
    \bm{M}^{\textrm{ab,b},+} =&\ \sqrt{W p} \bra{\textrm{bulk},+,\bm{p}} \bm{\sigma} \ket{\textrm{arc,b},\bm{p}_{\parallel}} = \sqrt{\frac{2\sin\theta\sin\phi}{1-\chi \sin\theta\cos\phi}}\frac{\cos\theta\pqty{-\bm{\alpha}_2 - i \hat{\bm{z}}}}{\chi + \sin\theta\cos\phi}  ,\\
    \bm{M}^{\textrm{ab,b},-} =&\ \sqrt{W p} \bra{\textrm{arc,b},\bm{p}_{\parallel}} \bm{\sigma} \ket{\textrm{bulk},-,\bm{p}}
    =\sqrt{\frac{2\sin\theta\sin\phi}{1 + \chi \sin\theta\cos\phi}}\frac{\cos\theta\pqty{\bm{\alpha}_2 - i \hat{\bm{z}}}}{\chi - \sin\theta\cos\phi}.
\end{align}
\end{subequations}
Letting $p_x \to - p_x$ in the second term of Eq. \eqref{eq:ab_current_bot_app}  the current may be written as
\begin{align}\label{eq:ab_current_bot_app_2}
    \bm{\Gamma}^{\textrm{ab,b}}_{ij} = \frac{16\eta\tau}{\tilde{W}} \int& \frac{\textrm{d}^3 p}{p}  \Theta(\cos\theta) \Theta(\sin\phi) \Theta(-\chi \cos\phi )   \frac{\delta \pqty{1 - p(1 - \chi \sin\theta\cos\phi )} \cos^2\theta\sin\theta\sin\phi}{\pqty{1-\chi \sin\theta\cos\phi}\pqty{1+\chi \sin\theta\cos\phi}^2}   
    \nonumber \\
    &\times \Bigg\{ \begin{pmatrix} \sin\theta\cos\phi - \chi \\ \sin\theta\sin \phi \end{pmatrix}
    (\bm{\alpha}_2 + i \hat{\bm{z}})_i (\bm{\alpha}_2 - i \hat{\bm{z}})_j 
    + \begin{pmatrix} -\sin\theta\cos\phi + \chi \\ \sin\theta\sin \phi \end{pmatrix} (\bm{\alpha}_2 - i \hat{\bm{z}})_i (\bm{\alpha}_2 + i \hat{\bm{z}})_j  \Bigg\}.
\end{align}
This may be evaluated straightforwardly. The resulting response tensor in the thick slab basis is
\begin{equation}
    \Gamma^{\textrm{ab,thick}}_x = \Gamma^{\textrm{ab,b}}_x = \frac{\eta\tau}{W}\ 
    \begin{pmatrix}
    0 & 0 & 0 \\ 0 & 0 & i \frac{16\pi}{3} \chi  \\ 0 & - i \frac{16\pi}{3}  \chi  & 0
    \end{pmatrix},\ 
    \Gamma^{\textrm{ab,thick}}_y = \Gamma^{\textrm{ab,b}}_y = \frac{\eta\tau}{W}\  
    \begin{pmatrix}
    0 & 0 & 0 \\ 0 & 2 \pi \ln 2 &  0  \\ 0 & 0  & 2 \pi  \ln 2
    \end{pmatrix}.
\end{equation}

The current due to the top arc (present only in the thin slab regime) 
may be obtained analogously. 
The result for the 
top arc in the basis
$\hat{\bm{x}}=\bm{\beta}_1$, 
$\hat{\bm{y}}=\bm{\beta}_2$ is 
\begin{align}
    \Gamma^{\textrm{ab,t}}_x = \frac{\eta\tau}{W}\ 
    \begin{pmatrix}
    0 & 0 & 0 \\ 0 & 0 & i \frac{16\pi}{3} \chi  \\ 0 & - i \frac{16\pi}{3}  \chi  & 0
    \end{pmatrix},\ \ \ \ 
    \Gamma^{\textrm{ab,t}}_y = -\frac{\eta\tau}{W}\  
    \begin{pmatrix}
    0 & 0 & 0 \\ 0 & 2 \pi  \ln 2 &  0  \\ 0 & 0  & 2 \pi  \ln 2
    \end{pmatrix}.
\end{align}
In the thin-slab case both 
contributions combine into the total 
arc-bulk contribution. Introducing the 
rotation operator 
\begin{equation}
    R(\varphi) = \begin{pmatrix}
    \cos\varphi & -\sin\varphi &0\\
    \sin\varphi & \cos\varphi&0\\
    0&0&1    \end{pmatrix},
\end{equation}
the total arc-bulk contribution, in the thin-slab basis [Fig.\ \ref{fig:directions}(c)], 
is given by 
\begin{align}
   \Gamma^{\textrm{ab,thin}}_{abc} =&\  
   R(\Delta)_{ai}R(\Delta)_{bj}R(\Delta)_{ck}\Gamma^{\textrm{ab,t}}_{ijk} + 
   R(-\Delta)_{ai}R(-\Delta)_{bj}R(-\Delta)_{ck}\Gamma^{\textrm{ab,b}}_{ijk},
\end{align}
which evaluates to
\begin{align}
 \frac{\bm{\Gamma}^{\textrm{ab,thin}}}{4\pi\eta\tau/W}  =&\  
 \hat{\bm{x}}\, \begin{pmatrix}
 \ln 2\,  \sin ^3\Delta\,  & 0 & 0 \\
 0 & \ln 2\,  \sin \Delta\,  \cos ^2\Delta\,  & i \frac{8}{3}   \chi \cos ^2\Delta\,  \\
 0 & -i \frac{8}{3}  \chi \cos ^2\Delta\,  &  \ln 2\,  \sin \Delta\,  \\
   \end{pmatrix}+\hat{\bm{y}}\, \sin\Delta 
  \begin{pmatrix}
 0 & \ln 2\,   \cos ^2\Delta\,  & -i \frac{8 }{3} \chi \sin \Delta\,  \\
 \ln 2\,  \cos ^2\Delta\,  & 0 & 0 \\
 i \frac{8}{3}   \chi \sin \Delta\,  & 0 & 0 
  \end{pmatrix}.
\end{align}

\section{Photogalvanic current due to bulk-bulk excitations}\label{app:calculation_bb}
This appendix section is structured as follows. First we perform some general manipulations. Next we discuss the thick and thin slab limits separately. In the thick slab limit, we may let
$W \to \infty$ and ignore the quantization condition Eq. \eqref{eq:bc_eq}. Corrections to the infinite system response arise from the spatial variation of the electromagnetic field corresponding to finite $\bm{k},\ 1/\delta$. These are of the same order of magnitude as the current due to arc-bulk excitations. Conversely, in the thin slab limit, the spatial dependence of the electromagnetic field may be ignored ($\bm{k} =0, 1/\delta = 0$), but the quantization condition due to finite $W$ leads to corrections of the same order as the arc-bulk current. 

We start from Eq. \eqref{eq:pge_formula_main} and specify to bulk states $m = (\bm{p},+)$ and $n = (\bm{q},-)$, \begin{align}\label{eq:bb_current_general}
     \bm{\Gamma}^{\textrm{bb}}_{ij}
     =&\ \frac{ 8\pi \eta\tau }{ \tilde{W}^3} \int \textrm{d}^2 \tilde{p}_{\parallel} \sum_{\tilde{p}_z \tilde{q}_z}  \pqty{\hat{\bm{v}}_{\bm{p}+} + \hat{\bm{v}}_{\bm{q}-}  } \delta\pqty{1 - \tilde{p} - \tilde{q}} 
     \bqty{\tilde{\bm{M}}^{\textrm{bb}}  \otimes \pqty{\tilde{\bm{M}}^{\textrm{bb}} }^*}_{ij}
     \Bigg\vert_{\bm{q}_{\parallel} = \bm{p}_{\parallel} - \bm{k}_{\parallel}}
\end{align} 
where we defined the dimensionless bulk-bulk matrix elements
\begin{equation}
    \tilde{\bm{M}}^{\textrm{bb}} = \tilde{W} \bra{\textrm{bulk},+,\tilde{\bm{p}}} \bm{\sigma} e^{i \tilde{z}  \tilde{k}_z -\tilde{z} /\tilde{\delta}} \ket{\textrm{bulk},-,\tilde{\bm{q}}}.
\end{equation}
Note that as opposed to the main text here we explicitly include the light momentum $\bm{k}$. Since $k=\omega/c$, the light momentum is a factor 
$v/c$ smaller than the typical momenta of excited 
states $\sim 1/\ell = \omega/v$.
We will show below that it does not contribute at the 
relevant order of magnitude, in agreement with the argumentation in the main text. We set $\bm{k} = (k_{\parallel} \cos\gamma,k_{\parallel} \sin\gamma,k_z)$ and keep terms to first order in $\tilde{k}_z,\, \tilde{k}_\parallel \sim v/c$. 

We proceed by performing the integral over $p_{\parallel}$ using the conservation of energy. The delta function gives the condition
\begin{align}\label{eq:conservation_of_energy_delta}
      g(p_{\parallel}) =&\ 1 - \sqrt{p_{\parallel}^2 + p_z^2} 
      - \sqrt{p_{\parallel}^2 + q_z^2 - 2p_{\parallel}k_{\parallel} \cos (\phi - \gamma) + k_{\parallel}^2} \nonumber = 0.
\end{align}
This has a real solution if, to leading order in $k_{\parallel}$,
\begin{align}
    p_z + q_z < 1 \textrm{ or } P < 1/2,
\end{align}
where we defined new variables
\begin{align}\label{eq:P_Q_change_of_vars}
    P =&\ \frac{1}{2}(p_z + q_z),\ Q = \frac{1}{2}(p_z - q_z)
\end{align}
with $P > 0,\ P > Q > -P$. The solution to $g(p_{\parallel}) = 0$ is given by
\begin{align}\label{eq:p_para}
    p_{\parallel} =&\ \frac{1}{2}\sqrt{(1-4P^2)(1-4Q^2)}  
    + \frac{k_{\parallel}}{2} \cos  (\phi - \gamma) \left(1+4PQ\right).
\end{align}
Performing the integration over $p_{\parallel}$ also gives rise to the Jacobian factor 
\begin{align}
    \frac{1}{\abs{g'(p_{\parallel})}} = \bqty{ \frac{p_{\parallel} + p_z (\textrm{d} p_z/\textrm{d} p_{\parallel}) }{p} + \frac{p_{\parallel} + q_z(\textrm{d} q_z/\textrm{d} p_{\parallel})- k_{\parallel} \cos (\phi - \gamma)}{q} }^{-1}.
\end{align}
The derivatives $\textrm{d} p_z/\textrm{d} p_{\parallel}$ may be obtained from the boundary condition Eq. \ref{eq:bc_eq_app}. They contribute at order $1/W$. 
Altogether, the bulk-bulk response tensor is now
\begin{align}\label{eq:bb_current_app_general}
     \Gamma^{\textrm{bb}}_{ijk} = \frac{ 8\pi \eta\tau }{ W^3} \int \textrm{d}\phi \sum_{p_z q_z}   \frac{ p_{\parallel}  }{\abs{g'(p_{\parallel})}} \pqty{\hat{\bm{v}}_{\bm{p}+} + \hat{\bm{v}}_{\bm{q}-}  }_i  M^{\textrm{bb}}_j \pqty{M^{\textrm{bb}}_k}^*  \Theta\bqty{1 - p_z - q_z} \Bigg\vert_{p_{\parallel}=p_{\parallel}(p_z,q_z,\phi),\ \bm{q}_{\parallel} = \bm{p}_{\parallel} - \bm{k}_{\parallel}}.  
\end{align}

Next, consider the matrix elements, $M^{\textrm{bb}}_i$. We can split them into a normalization factor that is common to all $M_i$ and a factor that depends on $i$,
\begin{equation}
    M^{\textrm{bb}}_i = W \bra{\textrm{bulk},+,\bm{p}} \sigma_i e^{ik_z z -  z/\delta} \ket{\textrm{bulk},-,\bm{q}} = \frac{1}{\sqrt{N_{\bm{p}+} N_{\bm{q}-}}} \mathcal{M}_i,
\end{equation}
where the $N_{\bm{p}\pm}$ are defined in Eq. \eqref{eq:normalization}. Using 
\begin{align}
    \bra{\alpha_+} \bm{\sigma} \ket{\alpha_+} =&\  - \bra{\alpha_-} \bm{\sigma} \ket{\alpha_-} = \bm{\alpha}_1,\  
    \bra{\alpha_+} \bm{\sigma} \ket{\alpha_-} = \bra{\alpha_-} \bm{\sigma} \ket{\alpha_+}^* = i \bm{\alpha}_2 + \hat{\bm{z}},
\end{align}
the $\mathcal{M}_i$ read 
\begin{align}\label{eq:bb_matrix_el_p_q}
    \bm{\mathcal{M}} =&\ 
        \Bqty{ 
            p_z q_z  I_1 
            + \bqty{
                p_2 q_2 +(p-\chi p_1 )(q + \chi  q_1 ) 
            } I_2 \nonumber 
            -  (p_z q_2 I_3 + p_2 q_z I_4) 
        }  \bm{\alpha}_1 \nonumber \\
        &
        + \Bqty{  
            \bqty{ \chi  \cdot (p q_2 - p_2 q) - (p_1 q_2 + p_2 q_1) } I_2
            +\pqty{p_z q_1 I_3 + p_1 q_z I_4}
            +\chi\pqty{p_z q I_3 - p q_z I_4}
       }\bm{\alpha}_2 \nonumber \\
        &
        + i\Bqty{ 
            \bqty{ \chi (p q_2 + p_2 q) - (p_1 q_2 - p_2 q_1) } I_2
            -\pqty{p_z q_1 I_3 - p_1 q_z I_4}
            -\chi\pqty{p_z q I_3 + p q_z I_4}
       }\hat{\bm{z}},
\end{align}
where we defined $x_i = \bm{\alpha}_i \cdot \bm{x}$ as well as the integrals
\begin{subequations}\label{eq:integrals}
 \begin{align}
    I_1 =&\ \int_0^W\textrm{d}z\ e^{(ik_z -1/\delta)z}\cos p_z z\cos q_z z ,\ \ \  I_2 = \int_0^W\textrm{d}z\ e^{(ik_z -1/\delta)z}\sin p_z z\sin q_z z, \\
    I_3 =&\ \int_0^W\textrm{d}z\ e^{(ik_z -1/\delta)z}\cos p_z z\sin q_z z ,\ \ \  I_4 = \int_0^W\textrm{d}z\ e^{(ik_z -1/\delta)z}\sin p_z z\cos q_z z.
\end{align}
\end{subequations}
These lead to conservation of ``momentum" perpendicular to the boundary if $W$ or $\delta$ become large. It will prove convenient to also define the combinations
\begin{align}
    I_\pm =&\ p_z I_3 \pm q_z I_4 = P(I_3 \pm I_4) + Q(I_3 \mp I_4), 
\end{align}
To leading order in $k_{\parallel}$, this gives  
\begin{align}
    p_z q I_3 \pm p q_z I_4 =&\ \frac{1}{2}\bqty{I_\pm - \pqty{4PQ + 2\bm{k}_{\parallel}\cdot\bm{p}_{\parallel}} I_\mp}.
\end{align}

\subsection{Thick slab $ W \gg \delta$}\label{app:calculation_bb_thick}
In this limit, dominant corrections are of 
order $1/\delta$. They stem from the 
light penetration depth 
$\delta=1$ and the finite light momentum $\bm{k}$
with $k\sim 1/\delta$. 
Ignoring the quantization of $p_z,q_z$, which give corrections of order 
$1/W \ll 1/\delta$ we replace 
\begin{align}
    \sum_{p_z q_z} \to  \int_0^{\infty} \textrm{d}p_z\int_0^{\infty} \textrm{d}q_z \to \frac{2 W^2}{\pi^2} \int_0^{\infty}\textrm{d}P \int_{-P}^P \textrm{d}Q.
\end{align}
We expand the bulk-bulk current in orders of $1/\delta$ and $k_i$. 

\subsubsection{Leading order bulk-bulk contributions $(\Gamma^{\textrm{bb,thick}} )$}
For the leading-order contributions we set $\bm{k}_{\parallel} = 0$  (i.e., $\bm{p}_{\parallel} = \bm{q}_{\parallel}$) and analyze integrals of the type 
\begin{align}\label{eq:I_iI_J_integrals}
    \int_0^{1/2}\textrm{d}P  \int_{-P}^P \textrm{d}Q\ I_i I_j^* f(P,Q), 
\end{align}
which enter the current formula
with a smooth kernel $f(P,Q)\sim \partial_{P,Q} f(P,Q)\sim 1$ ($f$ does not depend on the small parameter).
Considering $i=j=1$, we find
\begin{equation}
    \abs{I_1}^2 \simeq \frac{4 \left(\delta^{-2}+k_z^2\right) \left(P^2+Q^2\right)^2}{\left(\delta^{-2}+(k_z-2 P)^2\right) \left(\delta^{-2}+(k_z+2 P)^2\right) \left(\delta^{-2}+(k_z-2 Q)^2\right) \left(\delta^{-2}+(k_z+2 Q)^2\right)}.
\end{equation}
The leading contribution after integration
is the term proportional to $P^4$. 
Next, we use that   
\begin{align}
    \int_{-\infty}^{\infty}\textrm{d}Q \frac{4(k_z^2 + \delta^{-2})}{\left(\delta^{-2}+(k_z-2 Q)^2\right) \left(\delta^{-2}+(k_z+2 Q)^2\right)} = \pi \delta,
\end{align}
so that to leading order in $k_z$ and $1/\delta$ we can simplify
\begin{align}
    \frac{4(k_z^2 + \delta^{-2})}{\left(\delta^{-2}+(k_z-2 Q)^2\right) \left(\delta^{-2}+(k_z+2 Q)^2\right)} \to \pqty{\pi\delta} \delta(Q).
\end{align}
$\abs{I_2}^2$ and $I_1 I_2^*$ are 
similar to $\abs{I_1}^2$ with
 $\left(P^2+Q^2\right)^2$ replaced by $\left(P^2-Q^2\right)^2$ or $\left(P^4- Q^4\right)$, respectively. They clearly give the same leading order behaviour. For $P \gg v/c$ we have to leading order 
\begin{equation}
    \abs{I_1}^2 \simeq \abs{I_2}^2 \simeq I_1 I_2^*  \simeq \frac{\pi\delta}{16}  \delta(Q).
\end{equation}
Since $f(P,Q)$ is smooth, corrections from 
small $P\sim v/c$ are of higher order
in $v/c$.

Consider next $\abs{I_\pm}^2$  and the corresponding cross-term. It is
\begin{equation}
    \abs{I_+}^2 \simeq \frac{4 \left(\delta^{-2}+k_z^2\right)^2 \left(P^2+Q^2\right)^2}{\left(\delta^{-2}+(k_z-2 P)^2\right) \left(\delta^{-2}+(k_z+2 P)^2\right) \left(\delta^{-2}+(k_z-2 Q)^2\right) \left(\delta^{-2}+(k_z+2 Q)^2\right)} \simeq \frac{\pi \left(\delta^{-2}+k_z^2\right)\delta}{16}   \delta(Q).
\end{equation}
Due to the extra factor of $\delta^{-2}+k_z^2$ this may safely be ignored. However,
\begin{equation}
    \abs{I_-}^2 \simeq \frac{64 P^2 Q^2  \left(P^2-Q^2\right)^2}{\left(\delta^{-2}+(k_z-2 P)^2\right) \left(\delta^{-2}+(k_z+2 P)^2\right) \left(\delta^{-2}+(k_z-2 Q)^2\right) \left(\delta^{-2}+(k_z+2 Q)^2\right)}
\end{equation}
does not come with a small factor in the numerator at all. Naively one might expect a contribution of order $\delta^3$. The factor of $Q^2$ reduces this to a contribution of order $\delta$. The leading order contribution stems from the lowest power in $Q$, i.e. the term proportional to $P^6 Q^2$. Then, integrating by parts
\begin{align}
    \int_{-P}^P \textrm{d}Q &\ \frac{  64 Q^2  f(P,Q) }{\left(\delta^{-2}+(k_z-2 Q)^2\right) \left(\delta^{-2}+(k_z+2 Q)^2\right)} = - \frac{4\delta}{k_z} \int_{-P}^P \textrm{d}Q\ Q  f(P,Q) \partial_Q \bqty{ \frac{\pi}{2} - \arctan\left(\frac{\delta^{-2}-k_z^2+4 Q^2}{2 k_z /\delta}\right)} \nonumber \\
    &=\ 4\pi\delta\Bqty{ P\bqty{f(P,-P)-f(P,P)} \delta(P)  +  \int_{-P}^P \textrm{d}Q\ \bqty{f(P,Q) + Q \partial_Q f(P,Q)} \delta(Q) } \nonumber \\
    &=\ 4\pi\delta f(P,0)  
\end{align}
Here, we used  
\[
    \frac{1}{k_z} \int_{-\infty}^\infty \textrm{d}Q\ \bqty{ \frac{\pi}{2} - \arctan\left(\frac{\delta^{-2}-k_z^2+4 Q^2}{2 k_z /\delta }\right)} = \pi,
\]
s.t. the integrand is again a delta function for $k_z,1/\delta \to 0$. Hence, for $P \gg v/c$
\begin{equation}
     \abs{I_-}^2 \simeq \frac{\pi\delta}{4}P^2 \delta(Q).
\end{equation}
For the cross-term we find by a similar argument
\begin{equation}
    I_+ I_-^* \simeq -\frac{\pi (k_z+i/\delta)^2}{16 }\delta P \partial_Q \delta(Q),
\end{equation}
which is again negligible.

Finally, consider cross-terms of the type $I_1 I_+^*$, $I_2 I_+^*$, $I_1 I_-^*$, $I_2 I_-^*$. The former two  read to leading order
\begin{equation}
    I_i I_+^* \simeq \frac{4 (k_z^2 + \delta^{-2})(\delta^{-1}+i k_z) P^4  }{\left(\delta^{-2}+(k_z-2 P)^2\right) \left(\delta^{-2}+(k_z+2 P)^2\right) \left(\delta^{-2}+(k_z-2 Q)^2\right) \left(\delta^{-2}+(k_z+2 Q)^2\right)} \simeq \frac{\pi(\delta^{-1}+i k_z)}{16}\delta\ \delta(Q).
\end{equation}
The latter two read 
\begin{equation}
    I_i I_-^* \simeq \frac{-16 (1/\delta - ik_z) P^5 Q }{\left(\delta^{-2}+(k_z-2 P)^2\right) \left(\delta^{-2}+(k_z+2 P)^2\right) \left(\delta^{-2}+(k_z-2 Q)^2\right) \left(\delta^{-2}+(k_z+2 Q)^2\right)} \simeq \frac{\pi (k_z \delta-i)}{16 } P \partial_Q \delta(Q).
\end{equation}

With this we can now easily calculate the bulk-bulk current in the thick slab limit at leading order. We drop all combinations $I_i I_j^*$ with $i,j \in \{ 1,2,+,-\}$ which do not give a order-$\delta$ contribution. The remaining combinations all give a delta function $\delta(Q)$, corresponding to conservation of perpendicular momentum, $p_z = q_z$, and hence $p=q=1/2$ and $p_{\parallel} = \sqrt{1- 4P^2}/2 = \sqrt{1- p_z^2}/2 $. 
Identifying the combinations $I_i I_j^*$ giving rise to order-$\delta$ terms motivates the following definition:
\begin{align}\label{eq:Mexp}
    \bm{\mathcal{M}} \simeq \bm{\Sigma}_0  \mathcal{I}_0 + \bm{\Sigma}_1 \mathcal{I}_1,
\end{align}
where $\mathcal{I}_0 \simeq 2 p^2 I_1  \simeq 2 p^2 I_2$ and $\mathcal{I}_1 = p^2 I_- /P$. Here $\bm{\Sigma}_0$ is (up to normalization) the matrix element at fixed momentum,
\begin{align}\label{eq:sigma_0}
    \bm{\Sigma}_0  \equiv&\  \pqty{1-\sin^2\theta\cos^2\phi} \bm{\alpha}_1 -\sin^2\theta\cos\phi\sin\phi\ \bm{\alpha}_2
    +i \chi \sin\theta\sin\phi\ \hat{\bm{z}}
    \propto \bra{+,\bm{p}} \bm{\sigma} \ket{-,\bm{p}},
\end{align}
in terms of the spherical coordinates $p_z = p \cos\theta$, $p_{\parallel} = p \sin\theta$, while 
\begin{align}
    \bm{\Sigma}_1 \equiv&\    \chi \cos\theta\ \bm{\alpha}_2 \nonumber - i \cos\theta \sin\theta\cos\phi\ \hat{\bm{z}}.
\end{align}
This term in the matrix element only arises if one allows for $p_z \neq q_z$. It arises because $I_{3,4}$ may become large if $p_z - q_z \sim 1/\delta$ even though they vanish for $p_z = q_z$ (or $p_z - q_z \sim 1/W$ if the integral is cut off by the thickness of the slab, see below).
The normalization factor evaluates to 
\begin{align}
    \frac{1}{N_{\bm{p}+} N_{\bm{p}-}} = \frac{1}{p^4 (1-\sin^2\theta \cos^2\phi)}.
\end{align}
Altogether, after transforming $P =( \cos\theta )/2$, 
we have 
\begin{align}\label{eq:bb_current_thick_almost_final}
    \bm{\Gamma}^{\textrm{bb,thick}}_{0,ij} 
     = \frac{ \eta\tau \delta }{ W} \int \textrm{d}\Omega&\  \frac{\Theta(\cos\theta)\begin{pmatrix}
     \cos\phi \\ \sin\phi 
     \end{pmatrix}  \sin\theta }{1-\sin^2 \theta \cos^2\phi} 
     \Bqty{  \Sigma_{0,i}\Sigma_{0,j}^* + \Sigma_{1,i}\Sigma_{1,j}^*  }.
\end{align}
The angular integrals may be evaluated straightforwardly. This gives Eq. \eqref{eq:thickleading}. 
Note that the result is symmetric under rotations in the $x,y$-plane and thus independent of the direction of the boundary conditions. 

\subsubsection{Subleading order  bulk-bulk contributions $(\delta \Gamma^{\textrm{bb,thick}})$}

We now consider the leading corrections $\delta J^\mathrm{bb}$ to 
 $J^\mathrm{bb}$. From the above calculations, one can expect a correction of order 
\begin{align}
\frac{\delta \Gamma^{\textrm{bb}}}{ \Gamma^{\textrm{bb}}  } \sim \frac{v}{c} \ll 1. 
\end{align}
stemming from two different sources: first, from an in-plane momentum shift 
due to the finite light momentum $\bm{k}_{\parallel}$, and second, from the  
corrections to the integrals $I_i I_j^*$ of order $(1/\delta)^0,\ k_z^0$. 
Consider first corrections due to finite light momentum. Since these are already of the same magnitude as the corrections due to finite-size as well as the arc-bulk current, one may ignore the slab geometry here. 
It is straightforward to verify that for a bulk Weyl cone these corrections vanish. We thus expect that 
the relevant corrections due to a finite $\bm{k}$ vanish also in the slab. 
We first consider  finite $k_{\parallel}$. The products 
$I_i I^*_j$  do not involve $k_{\parallel}$ and are thus approximated as in the leading-order calculation above. In $\bm{\mathcal{M}}$ we 
can thus again separate out 
$\mathcal{I}_0 \simeq 2 p^2 I_1  \simeq 2 p^2 I_2$ and $\mathcal{I}_1 = p^2 I_- /P$  and expand the prefactors up to leading order in $k_\parallel$,
\begin{align}
    \bm{\mathcal{M}} \simeq \bm{\Sigma}'_0  \mathcal{I}_0 + \bm{\Sigma}'_1 \mathcal{I}_1, \textrm{ where } \bm{\Sigma}'_i = \bm{\Sigma}_i + k_{\parallel} \delta \bm{\Sigma}_i + \order{k_{\parallel}^2}, 
\end{align}
where, introducing the shorthand notation $c_x = \cos x, s_x = \sin x$, 
\begin{subequations}
\begin{align}
     \delta\bm{\Sigma}_0 
    =&\
    \frac{1}{2}\bqty{ \pqty{\chi s_{\theta}^2 c_{\phi}^2   + 2 s_{\theta} c_{\phi}  s_{\phi}^2 -\chi }c_{\gamma} + c_{\phi} s_{\phi} \pqty{ \chi s_{\theta}^2 -  2 s_{\theta} c_{\phi}  }s_{\gamma}   } \bm{\alpha}_1 \nonumber \\
    &+ \frac{1}{2}\bqty{
    c_{\phi} s_{\phi}  \pqty{\chi s_{\theta}^2  - 2 s_{\theta} c_{\phi}  }c_{\gamma}
    + \pqty{
    \chi s_{\theta}^2 s_{\phi}^2 - 
    2 s_{\theta} c_{\phi} s_{\phi}^2 - \chi  }s_{\gamma}
    + s_{\theta} s_{\phi+\gamma}
       } \bm{\alpha}_2 \nonumber \\
    &-\frac{i}{2} s_{\gamma-\phi} \bqty{ s_{\theta} -  \chi c_{\phi} } \hat{\bm{z}},  \\
    \delta\bm{\Sigma}_1 
    =&\
    \frac{1}{2}\Bqty{ s_{\gamma} \bm{\alpha}_1  - c_{\gamma} \bm{\alpha}_2 + i \bqty{ -\pqty{s^2_{\phi} + \chi s_{\theta} c_{\phi} }c_{\gamma} + \pqty{s_{\phi} c_{\phi} -\chi s_{\theta} s_{\phi} }s_{\gamma} } \hat{\bm{z}} }.
\end{align}
\end{subequations}
Other terms entering the current formula expand as
\begin{subequations}
\begin{align} \label{eq:velocity_jacobian_factor}
     \frac{ p_{\parallel}    }{\abs{g'(p_{\parallel})}}  
    \pqty{ \hat{\bm{v}}_{\bm{p}+} + \hat{\bm{v}}_{\bm{q}-}}
    =&\  
    \frac{s_\theta}{2}
    \begin{pmatrix}
    c_\phi \\ s_\phi 
    \end{pmatrix}
    +
    \frac{k_{\parallel} }{2} 
    \begin{pmatrix}
     c_{\gamma - 2\phi} \\ -s_{\gamma-2\phi} 
    \end{pmatrix}
    + \order{k_{\parallel}^2},\\
    N_{\bm{p}+} N_{\bm{p}-} =&\  \frac{1}{16} \pqty{1-s^2_{\theta} c^2_{\phi}} + \frac{ k_{\parallel}}{8} \bqty{\pqty{s_\theta c_\phi - \chi} c_\gamma + \pqty{ \chi s_{\theta} - c_{\phi} }s_\theta c_\phi c_{\gamma - \phi}} 
    + \order{k_{\parallel}^2}.
\end{align}
\end{subequations}
We now specify to the basis of the thick slab, $\hat{\bm{x}} = \bm{\alpha}_1, \hat{\bm{y}} = \bm{\alpha}_2$. 
For $\gamma = 0$ (i.e., $\bm{k}_\parallel = k_\parallel \hat{\bm{x}}$) 
this combines to the total correction 
of the response tensor
\begin{subequations}
\begin{align}
\delta \Gamma_{x}^{\textrm{bb},\bm{k}_{\parallel}}  \propto&\  
\int_0^{\frac{\pi}{2}} \textrm{d}\theta
\int_0^{2\pi} \textrm{d}\phi
   \left(
\begin{array}{ccc}
  c_{\theta}^2-4 s_{\theta }^2s_{\phi }^4+\left(5 s_{\theta }^2- 2\right) s_{\phi }^2
 & 2  s_{\theta }^2  c_{\phi }  s_{\phi } \left(2 s_{\phi }^2-1\right) 
 &  i s_{\theta } s_{\phi } \left(3 s_{\phi }^2-2\right) 
 \\
 2  s_{\theta }^2  c_{\phi }  s_{\phi } \left(2 s_{\phi }^2-1\right) 
 & 1 + 4 s_{\theta }^2 s_{\phi }^4 +\left(3 c_{\theta }^2-5\right) s_{\phi }^2 
 &  i c_{\phi } s_{\theta } \left(1-3 s_{\phi }^2\right) 
 \\
 - i s_{\phi } s_{\theta }  \left(3 s_{\phi }^2 - 2\right) 
 & - i c_{\phi } s_{\theta } \left(1-3 s_{\phi }^2\right) 
 & -s_{\theta }^2 \left(2 s_{\phi }^2-1\right) \\
\end{array}
\right) , \\ 
\delta \Gamma_{y}^{\textrm{bb},\bm{k}_{\parallel}} \propto&\  
\int_0^{\frac{\pi}{2}} \textrm{d}\theta
\int_0^{2\pi} \textrm{d}\phi
   \left(
\begin{array}{ccc}
 c_{\phi } s_{\phi } \left(2 c_{\theta }^2 +4  s_{\theta }^2 s_{\phi }^2 \right) 
 &  s_{\theta }^2 s_{\phi }^2 \left(4 s_{\phi }^2-3\right) 
 & -3 i c_{\phi } s_{\theta } s_{\phi }^2 
 \\
 s_{\theta }^2 s_{\phi }^2 \left(4 s_{\phi }^2-3\right)
 & c_{\phi } s_{\phi } \left(2 - 4  s_{\theta }^2 s_{\phi }^2\right) 
 &  i s_{\theta } s_{\phi } \left(2-3 s_{\phi }^2\right) 
 \\
 3 i c_{\phi } s_{\theta } s_{\phi }^2 
 &  i s_{\theta } s_{\phi } \left(3 s_{\phi }^2-2\right) 
 & 2 s_{\theta}^2 c_{\phi }  s_{\phi } \\
\end{array}
\right).
\end{align}
\end{subequations}
It is straightforward to confirm that these expressions vanish upon integration over $\phi$. Similar expressions for $\gamma = \pi/2$ 
(i.e., $\bm{k}_\parallel = k_\parallel \hat{\bm{y}}$) also vanish. 

The remaining corrections are corrections to the products 
$I_iI_j$, which have been discussed above, 
of order $(k_z\pm i /\delta)^0$. We verified numerically, that the 
correction due to a finite $k_z$ vanishes as expected.  
Fitting the numerically evaluated response 
tensor (for $\delta \in \{10^2 , 2\times 10^{2} , ... ,10^{3}\}$) 
to an expansion up to second order in $1/\delta$ 
we find (rounding to the second decimal)
\begin{subequations}
\begin{align}
\delta \Gamma^{\textrm{bb,thick}}_x       =&\ \frac{\eta\tau}{\tilde{W}}    \left(
\begin{array}{ccc}
 0 & 4.19 & 0 \\
 4.19 & 0 &  -16.75  i \chi \\
 0 &  16.75   i \chi& 0 \\
\end{array}
\right), \\
\delta \Gamma^{\textrm{bb,thick}}_y   =&\  \frac{\eta\tau}{\tilde{W}}
 \left(
\begin{array}{ccc}
 -4.19 & 0 &  9.87 i \chi \\
 0 & -4.20  & 0 \\
 -9.87 i \chi & 0 & -8.40 \\
\end{array}
\right) .
\end{align}
\end{subequations}
To estimate the accuracy of these results we compare the numerical values given here to those obtained by fitting expansions to higher order in $1/\delta$ as well as by adding/removing data points corresponding to the smallest values of $\delta$. These changes in the fitting procedure lead to changes in the numerical coefficients of $\leq 0.5\%$. The error analysis is summarized in Table \ref{tab:error_tab_2}.

\begin{table*} 
\begin{tabular}{ p{1.5cm}|p{2cm}|p{2cm}|p{2cm}|p{2cm}|p{2cm}|p{2cm}  }
& $\delta \Gamma^{\textrm{bb}}_{xxy}$ & $\delta \Gamma^{\textrm{bb}}_{xyz}$ & $\delta \Gamma^{\textrm{bb}}_{yxx}$ & $\delta \Gamma^{\textrm{bb}}_{yyy}$
        & $\delta \Gamma^{\textrm{bb}}_{yzz}$ & $\delta \Gamma^{\textrm{bb}}_{yzx}$ \\
\hline       
Error   
        & $ 5 \times 10^{-4}$ & $ 3 \times 10^{-3} $ 
        & $ 5 \times 10^{-5} $ & $ 5 \times 10^{-3} $
        & $ 1 \times 10^{-3} $ & $ 1 \times 10^{-5} $ \\
\end{tabular}
\caption{\label{tab:error_tab_2}
Estimated inaccuracy of numerical results for the subleading bulk-bulk response tensor $\delta \Gamma^{\textrm{bb}}$. The error corresponds to the statistical relative standard error of a fit with first order polynomial in $1/\delta$ and $\delta \in \{10^{2.8},10^{3.0},...,10^{3.4}\}$ (for larger values of $\delta$ the integration is no longer stable due to the sharply peaked nature of the integrals $I_i$).  Use of higher order polynomials or inclusion of smaller $\delta$-points gives comparable results and deviations. } 
\label{length} 
\end{table*}

\subsection{Thin slab $ \delta \gg W \gg \ell $}\label{app:calculation_bb_thin}

In this limit, the leading finite-size corrections are $\sim 1/W$, 
as discussed in the main text. Corrections due to the 
spatial variation of the external field, which 
we found to give corrections of order 
$\sim 1/\delta$, are thus negligible and we can
set $\bm{k} = 0,\ \delta \to \infty$. We now work in the 
basis of the thin slab.

\subsubsection{Leading order ($\Gamma^{\textrm{bb,thin}}   $)}
To calculate the leading order response we 
disregard the quantization of $p_z,q_z$. Considering leading-order terms of the 
integral products $I_iI_j$, the dominant contributions read
\begin{equation}
    I_1^2 \simeq I_2^2 \simeq I_1 I_2 \simeq \frac{\pi \tilde{W}}{8} \delta(Q),\ I_-^2 \simeq \frac{\pi \tilde{W} P^2}{2} \delta(Q),
\end{equation}
all other combinations contribute only at higher order.
Since the difference to the leading contribution of the 
thick-slab case is in the constant prefactor, the 
leading-order bulk-bulk contribution in the thin slab limit
is given by the thick-slab result replacing $\delta/2 \to W$. The final response tensor reads
\begin{align}\label{eq:bb_thin_leading_order_app}
    \Gamma_{ijk}^{\textrm{bb}} =&\ i\chi  \frac{4\pi \eta\tau }{3} \varepsilon_{ijk}
    (1-\delta_{i,z}). 
\end{align}

\subsubsection{Subleading order ($\delta \Gamma^{\textrm{bb,thin}}$)}
The leading corrections to the bulk-bulk contribution 
in the thin slab limit are of order $1/W$. They can stem 
from the quantization of $p_z,q_z$, the associated 
corrections to the velocity in Eq. \eqref{eq:velocity_general}, and the
corrections $\delta N_{\bm{p}\pm}$ to the wavefunction normalization. The current is given by Eq. \eqref{eq:bb_current_app_general}
with $p_z,q_z$ solutions of
\begin{subequations}\label{eq:boundary_condition_equation_coupled_app_intermediate}
\begin{align} 
    \sin \Delta  =&\ \frac{\tan (p_z W)}{p_z} \Bigg[  p_{\parallel} \cos\phi - \chi p \cos \Delta \Bigg],\\
    \sin \Delta =&\ \frac{\tan (q_z W)}{q_z} \Bigg[  p_{\parallel} \cos\phi + \chi q \cos \Delta \Bigg],
\end{align}
\end{subequations}
where we defined the characteristic angle 
\begin{equation}
    \Delta = \frac{\beta - \alpha}{2} \in \bqty{-\pi,\pi}.
\end{equation}
These expressions as well as the tensors below are in the basis of the thin slab. 
Note that energy conservation makes 
$p_{\parallel}$ and thus also $p$ and $q$ depend on $(p_z,q_z)$. To evaluate the expression for the current for a given value of $\Delta$ we resort to numerics. We then attempt to extract the functional dependence of the nonzero tensor components by fitting appropriate polynomials in $\sin\Delta$ and $\cos\Delta$. 

We briefly outline the numerical strategy employed to extract the response tensor. For $\phi$-integration at fixed $W$ we employ standard numerical techniques relying on evaluation of the integrand for a discrete set of $\phi$-points. For each $\phi$-point we numerically determine all solutions to Eqs. \eqref{eq:boundary_condition_equation_coupled_app_intermediate} in the region $p_z + q_z < 1$. 
To determine $\delta\Gamma^{\textrm{bb}}$ we evaluate Eq. \eqref{eq:bb_current_app_general} at $W \in \{100,150,200,250\}$ and subtract the leading order term, Eq. \eqref{eq:bb_thin_leading_order_app}. We then fit to an expansion up to second order in $1/W$ and extract the coefficient of the $1/W$ term. In this way we determine all symmmetry-allowed elements of $\delta\Gamma^{\textrm{bb}}$ for 30 values of $\Delta \in [0,\pi/2]$ (the intervals $[-\pi,0]$ and $[\pi/2,\pi]$ may  be obtained from symmetry considerations, see Sec. \ref{app:symmetry}). Finally, we fit the components of the symmetric part of the response tensor to each element to an appropriate expansion in Fourier modes,
\[
    \sum_{n \textrm{ odd}}^{n_{\textrm{max}}} a^{\textrm{s}}_n  \sin (n \Delta),
\]
where we exploit that they are odd under $\Delta \to -\Delta$ and $\Delta \to \Delta + \pi$. Similarly, we fit the components of the anti-symmetric part to 
\[
    a^{\textrm{as}}_0 + \sum_{n \textrm{ even}}^{n_{\textrm{max}}} a^{\textrm{as}}_n  \cos (n \Delta),
\]
where we use that they are even under the above transformations. We found that $n_{\textrm{max}} = 3$ gives sufficiently good results with higher order coefficients satisfying $a_{n>3}/(\textrm{max } a_{n\leq 3}) \lesssim 10^{-3}$. Rounding to $10^{-2}$, the  subleading order response tensor due to bulk-bulk excitations in the thin slab limit is
\begin{subequations}\label{eq:response_tensor_thin_app}
\begin{align}
     \delta\Gamma^{\textrm{bb,thin}}_x =&\ \frac{\eta\tau}{W} \left(
\begin{array}{ccc}
 -14.14 \sin \Delta + 4.71 \sin 3 \Delta & 0 & 0 \\
 0 & -21.47 \sin \Delta -4.71 \sin 3\Delta  & -i\chi(26.71+ 6.88 \cos 2\Delta ) \\
 0 & i\chi(26.71+ 6.88 \cos 2 \Delta ) & -23.04 \sin\Delta \\
\end{array}
\right),\\
     \delta\Gamma^{\textrm{bb,thin}}_y =&\ \frac{\eta\tau}{W} \left(
\begin{array}{ccc}
 0 & 3.67 \sin \Delta -4.71 \sin 3\Delta & i\chi( 26.50 -  6.89 \cos 2\Delta ) \\
 3.67 \sin \Delta -4.71 \sin 3\Delta & 0 & 0 \\
 -i\chi( 26.50 -  6.89 \cos 2\Delta ) & 0 & 0 \\
\end{array}
\right).
\end{align} 
\end{subequations}
These results are accurate to the first decimal: the error estimates of the numerical integration scheme are on the order of $10^{-1}$ to $10^{-2}$. 
Similarly, altering the fitting procedure (e.g. by fitting to a first order expansion in $1/W$ or by removing data-points in $W$) leads to changes in the numerical coefficients on the order of roughly $10^{-2}$ with the largest changes, at about $1\%$, observed for the $\Delta$-independent terms in the circular components.
Note that averaging over $\Delta \in [-\pi,\pi]$ restores rotational symmetry around the $z$-axis, which implies that the $\Delta$-independent terms of $\Gamma_{xyz}$ and $\Gamma_{yzx}$ should actually be equal. Here, they differ by roughly $0.5 \%$, 
consistent with our error estimate.

\section{Lattice simulation of a thin slab}\label{app:ultrathin}

In this section we perform numerical calculations of the PGE response tensor 
for a lattice model of the Weyl slab in the thin-slab limit. It will extend the 
above semianalytical calculations, 
considered under the simplifying assumption $W\gg 1$ (the tilde is here 
still suppressed so that $W$ is in units of $\ell$), to the case of arbitrary $W$.
In the \emph{ultrathin} limit $W\sim 1$ the 
confinement-induced and bulk contributions are of the same order.

We consider the ultrathin limit using the lattice Hamiltonian, 
\begin{equation}\label{eq:latticeHam}
    H_{ij} = \frac{v}{2}\bqty{\sigma_x p_x +\sigma_y p_y
    + i \sigma_z  \delta_{i,j+1}   + \sigma_x\pqty{1- \delta_{i,j+1}}} + h.c.,
\end{equation}
where $i,j$ denote the site index at fixed in-plane momenta $p_x,p_y$. This lattice version of the original infinite-system Weyl Hamiltonian (lattice constant set to one) has been constructed replacing
\begin{equation}\label{eq:latticereg}
    \sigma_z p_z \mapsto \sigma_z \sin p_z + 
    \sigma_x \pqty{\cos p_z-1},
\end{equation}
such that the Hamiltonians coincide at small $p_z$ (the second term removes a spurious
Weyl cone at $p_z=\pm\pi$). Transformation into the site basis replaces
\begin{equation}
    \sin p_z \mapsto i\tfrac{1}{2}\pqty{
    \delta_{i,j+1}-\delta_{i,j-1}},\ \ \ \ \ \ \ 
     \cos p_z \mapsto \tfrac{1}{2}\pqty{
    \delta_{i,j+1}+\delta_{i,j-1}},
\end{equation}
which leads to \eqref{eq:latticeHam}. The Hamiltonian
\eqref{eq:latticeHam} can 
be considered for a finite site number. The choice of the Pauli 
matrix $\sigma_x$ for the 
second term in \eqref{eq:latticereg}
sets the  direction of the Fermi arc such that $\bm{\alpha}_2=\hat{\bm{x}}=-\bm{\beta}_2$, 
corresponding to $\Delta=\pi/2$ of the thin-slab case considered above
\cite{bovenzi_twisted_2018}.

\begin{figure}
    \centering
    \includegraphics[width=\textwidth]{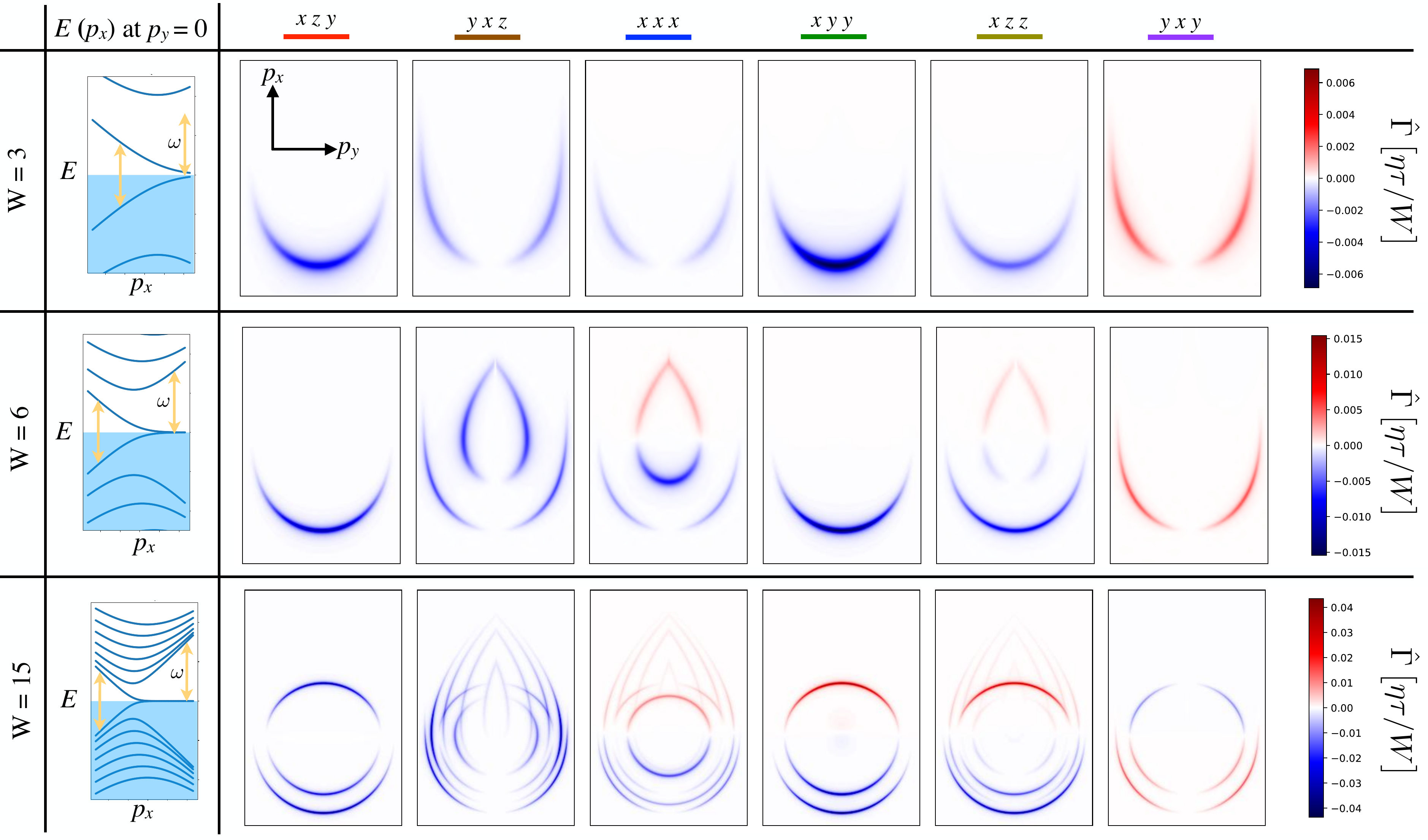}
    \caption{Parallel-momentum $(p_x,p_y)$ resolved response tensor $\hat{\Gamma}$ (    $\Gamma=\sum_{p_x,p_y}\hat{\Gamma}$) at three 
    widths $W=3,6,15$ for $v=1$ and 
    $\ell=5$ sites.  Left column indicates the 
    dispersion at $p_y=0$. }
    \label{fig:ultrathin}
\end{figure}

Numerical results for the PGE response tensor 
in Eq.  \eqref{eq:pge_formula_main} are obtained
via discretizing parallel momenta, numerically 
diagonalizing the Hamiltonian \eqref{eq:latticeHam},
and summing over all pairs of states (one 
below and one above the Fermi level). The 
numerical 
discretization spacing 
and numerical broadening of the 
delta-function expressing energy
conservation have been decreased until 
convergence of the results. 
Figure \ref{fig:lattice_results} shows the 
results for the response tensor as a function of the width. Fig. \ref{fig:ultrathin}
shows the contributions resolved in the in-plane momentum. One can clearly see the cusp-like lines of the arc-bulk excitations and the circular lines 
of bulk-bulk excitations, c.f.\ Fig.\ \ref{fig:arc_bulk_transitions}. The signs of the contributions and the presence/absence of arc-bulk contributions is as discussed in the main text.

\end{document}